\documentclass[twocolumn]{aastex63}

\usepackage{multirow}

\newcommand{\xmm} {{\it XMM-Newton}}
\newcommand{\chandra} {{\it Chandra}}
\newcommand{\nustar} {{\it NuSTAR}}
\newcommand{\swift} {{\it Swift}}
\newcommand{\suzaku} {{\it Suzaku}}
\newcommand{\erosita} {eROSITA}
\newcommand{\rosat} {ROSAT}
\newcommand{\swiftxrt} {{\it Swift}/XRT}
\newcommand{\swiftuvot} {{\it Swift}/UVOT}
\newcommand{\gaia} {{\it Gaia}}
\newcommand{\hst} {{\it HST}}

\newcommand{\cmsq} {cm$^{-2}$}
\newcommand{\nh} {$N_{\rm{H}}$}
\newcommand{\lx} {$L_{\rm{X}}$}

\newcommand{\degree}{{$^\circ$}}
\newcommand{\ergs}{\mbox{\thinspace erg\thinspace s$^{-1}$}}
\newcommand{\ergcms}{\mbox{\thinspace erg\thinspace cm$^{-2}$\thinspace s$^{-1}$}}

\newcommand{\cntrt}{counts\,s$^{-1}$}

\newcommand{\mbh} {$M_{\rm BH}$}

\newcommand{\msol} {$M_{\odot}$}

\newcommand{\srci} {Swift J130456.1-493158}
\newcommand{\srcii} {Swift J130511.5-492933}
\newcommand{\srciii} {2SXPS J235825.7-323609}
\newcommand{\srciv} {Swift J095520.7+690401}
\newcommand{\srcv} {Swift J235749.9-323526}

\shorttitle{Ultraluminous X-ray Transients}
\shortauthors{Brightman et al.}

\begin{document}

\title{A new sample of transient ultraluminous X-ray sources serendipitously discovered by \swiftxrt}

\author{Murray Brightman}
\affiliation{Cahill Center for Astrophysics, California Institute of Technology, 1216 East California Boulevard, Pasadena, CA 91125, USA}

\author{Jean-Marie Hameury}
\affiliation{Observatoire Astronomique de Strasbourg, CNRS UMR 7550, 67000 Strasbourg, France}

\author{Jean-Pierre Lasota}
\affiliation{Institut d'Astrophysique de Paris, CNRS et Sorbonne Universit\'es, UMR 7095, 98bis Boulevard Arago, 75014 Paris, France}
\affiliation{Nicolaus Copernicus Astronomical Center, Polish Academy of Sciences, Bartycka 18, 00-716 Warsaw, Poland}

\author{Ranieri D. Baldi}
\affiliation{INAF - Istituto di Radioastronomia, Via Piero Gobetti 101, I-40129 Bologna, Italy}

\author{Gabriele Bruni}
\affiliation{INAF - Istituto di Astrofisica e Planetologia Spaziali, Via del Fosso del Cavaliere 100, I-00133 Roma, Italy}

\author{Jenna M. Cann}
\affiliation{NASA Goddard Space Flight Center, 8800 Greenbelt Rd, Greenbelt, MD 20771 USA}

\author{Hannah Earnshaw}
\affiliation{Cahill Center for Astrophysics, California Institute of Technology, 1216 East California Boulevard, Pasadena, CA 91125, USA}

\author{Felix F\"{u}rst}
\affiliation{Quasar Science Resources SL for ESA, European Space Astronomy Centre (ESAC), Science Operations Departement, 28692 Villanueva de la Ca\~{n}ada, Madrid, Spain}

\author{Marianne Heida}
\affiliation{European Southern Observatory, Karl-Schwarzschild-Str 2, 85748 Garching bei M\"{u}nchen, Germany}

\author{Amruta Jaodand}
\affiliation{Cahill Center for Astrophysics, California Institute of Technology, 1216 East California Boulevard, Pasadena, CA 91125, USA}

\author{Margaret Lazzarini}
\affiliation{Cahill Center for Astrophysics, California Institute of Technology, 1216 East California Boulevard, Pasadena, CA 91125, USA}

\author{Matthew J. Middleton}
\affiliation{Department of Physics \& Astronomy, University of Southampton, Southampton, SO17 1BJ, UK}

\author{Dominic J. Walton}
\affiliation{Centre for Astrophysics Research, University of Hertfordshire, College Lane, Hatfield AL10 9AB, UK}

\author{Kimberly A. Weaver}
\affiliation{NASA Goddard Space Flight Center, 8800 Greenbelt Rd, Greenbelt, MD 20771 USA}

\email{murray@srl.caltech.edu}

\begin{abstract}
Ultraluminous X-ray sources (ULXs) are our best laboratories for studying extreme super-Eddington accretion. Most studies of these objects are of relatively persistent sources, however there is growing evidence to suggest a large fraction of these sources are transient. Here we present a sample of five newly reported transient ULXs in the galaxies NGC 4945, NGC 7793 and M81 serendipitously discovered in \swiftxrt\ observations. Swift monitoring of these sources have provided well sampled lightcurves, allowing for us to model the lightcurves with the disk instability model of \cite{hameury20} which implies durations of 60--400 days and that the mass accretion rate through the disk is close to or greater than the Eddington rate. Of the three source regions with prior {\it HST} imaging, color magnitude diagrams of the potential stellar counterparts show  varying ages of the possible stellar counterparts. Our estimation of the rates of these sources in these three galaxies is 0.4--1.3 year$^{-1}$.  We find that while persistent ULXs dominate the high end of galaxy luminosity functions, the number of systems that produce ULX luminosities are likely dominated by transient sources.

\end{abstract}

\keywords{}

\section{Introduction}

Ultraluminous X-ray sources (ULXs) are powerful X-ray sources found outside the nucleus of galaxies (see \cite{kaaret17}, \cite{fabrika21} and \cite{king23} for recent reviews). They exhibit luminosities in excess of $10^{39}$ \ergs\ which is the Eddington limit of the typical 10 \msol\ black hole found in our Galaxy. First identified in the early 1980s by the {\it Einstein Observatory} \citep{giacconi79}, the first fully imaging X-ray telescope put into space, they were originally thought to be more massive black holes, potentially intermediate-mass black holes \citep[\mbh$=100$--$10^5$ \msol, e.g.][]{colbert99}. However, more recently, consensus has shifted to view these sources as lower-mass super-Eddington accretors \citep[e.g.][]{middleton15a}. This was famously confirmed for some sources by the detection of pulsations, revealing their central engines to be neutron stars \citep[NSs, e.g.][]{bachetti14,fuerst16,israel17,israel17a} and not black holes at all. NSs have masses of only 1--2 \msol, implying their luminosities when assuming isotropic emission to be 100\,s of times the Eddington limit. ULXs are thus our best laboratories for studying extreme super-Eddington accretion. 

The vast majority of ULX studies have been on relatively persistent sources, i.e. sources that while some may be highly variable, are consistently active and have been detected by X-ray instruments for decades. Indeed there is evidence to suggest they have been active for much longer from the collisionally ionized bubbles surrounding sources such as Holmberg IX X-1, NGC 1313 X-2, NGC 7793 S26 and NGC 5585 ULX which have estimated dynamical ages of $\sim10^{5}$ years \citep{pakull02,pakull10,moon11,weng14,berghea20,soria21}. Studies of persistent ULXs have revealed their multicomponent X-ray spectra \citep[e.g.][]{gladstone09,walton18c}, coherent pulsations \citep[e.g.][]{bachetti14,fuerst16,israel17a,israel17}, ultrafast outflows \citep[e.g.][]{pinto16,kosec18}, super-orbital periods \citep[e.g.][]{walton16b,hu17,brightman19,brightman20}, cyclotron lines \citep{brightman18,walton18b}, among many other things.

However, in addition to persistent ULXs there are several known transient ULXs. Indeed one of these occurred in our own Galaxy, Swift~J0243.6+6124 \citep{cenko17,wilsonhodge18}, and another in the SMC, RX~J0209.6-7427 \citep{chandra20,vasilopoulos20b}. Both of these were found to be powered by NS accretors with a Be star companion. Type I Be X-ray binary outbursts occur when a neutron star, often in a wide eccentric orbit, accretes material as it passes through the decretion disk of its Be star companion \citep{reig11}. Type II outbursts are brighter and often reach the Eddington limit, as was the case with Swift~J0243.6+6124 and RX~J0209.6-7427. It is not clear if all transient ULXs are Be X-ray binaries, however M51 XT-1 \citep{brightman20} would be a candidate for a non-Be X-ray binary since it peaked at an X-ray luminosity of $10^{40}$ \ergs, much greater than seen in Be X-ray binaries.

Transient ULXs are far less well studied than their persistent counterparts, potentially skewing our understanding of super-Eddington accretion and of ULXs in general \citep{dage21}. This is mostly due to the lack of wide-field X-ray surveys with the sensitivity to detect these mostly extragalactic sources. \erosita\ was launched in 2019 and the data from its all sky surveys will have the potential to change this. Most ULXs known today have been identified serendipitously in pointed imaging X-ray observations by \xmm, \chandra\ and \swift\ \citep[e.g.][]{liu05,liu05b,winter06,swartz11,walton11,earnshaw19,kovlakas20}, with the latest catalog of ULX candidates containing 1843 sources \citep{walton21}. However, the relative rates of persistent and transient sources is unknown. A few detailed studies of transient ULXs discovered serendipitously have been presented in the literature \citep[e.g.][]{strickland01,soria07,middleton12,soria12,middleton13,carpano18,pintore18,liu19,vanhaaften19,earnshaw19b,brightman20,earnshaw20,walton21a,dage21,robba22}, however, a systematic search for transient ULXs is lacking.

NASA's {\it Neil Gehrels Swift Observatory} \citep[hereafter \swift,][]{gehrels04} observes 10\,s of targets a day, many of which are monitoring observations, with the data being quickly downloaded and made public. This allows for a near real time search for transients, and detailed follow up. We have already reported on the discovery of a tidal disruption event found this way \citep{brightman21}, and the \swift\ team have recently presented the Living Swift-XRT Point Source catalogue (LSXPS) and real-time transient detector \citep{evans22}. Here we report on results on transient ULXs from our own systematic search for X-ray transients in \swiftxrt\ observations. 

\section{The search for new X-ray transients}
\label{sec_srcs}

Beginning in $\sim$2019 October, we have routinely dowloaded a selection of new \swiftxrt\ observations on a $\sim$daily basis. Not all observations were downloaded due to time constraints. We searched for sources in these observations using the {\tt detect} function of the {\sc heasoft} tool {\sc ximage} and a signal to noise threshold of 3. The positions of the detected X-ray sources were then cross-correlated with latest versions of the \swift\ Point Source Catalog \citep[2SXPS,][]{evans20}, the Fourth \xmm\ Serendipitous Source Catalogue \citep[4XMM,][]{webb20}, the \chandra\ Source Catalog \citep[CSC2,][]{evans10} and the Second \rosat\ All-Sky Survey Source Catalogue \citep[2RXS,][]{boller16}. When the new \swift\ source was found to have no close counterpart in these catalogs, we first assessed if this is because the source position was not previously observed by an imaging X-ray telescope, or it was a genuine new source. If it appeared to be a new source, we investigated further by using the online tool provided by the University of Leicester\footnote{https://www.swift.ac.uk/user\_objects/} \citep{evans07,evans09} to determine the best position, and generate a lightcurve and spectrum of the source. All products from this tool are fully calibrated and corrected for effects such as pile-up and the bad columns on the CCD. All spectra were grouped with a minimum of one count per bin using the {\sc heasoft} v 6.28 tool {\tt grppha} and fitted in {\sc xspec} v 12.11.1 \citep{arnaud96}. The C statistic was used for fitting to source spectra with the background subtracted \citep{cash79}. Since the C statistic cannot formally be used when the background is subtracted, {\sc xspec} uses a modified version of the C statistic known as the W statistic to account for this. We describe the five new sources we found below.
 
\subsection{\srci, an X-ray transient in the field of NGC 4945}
\label{sec_J1304}

\begin{figure*}
\begin{center}
\includegraphics[trim=10 20 20 0, width=180mm]{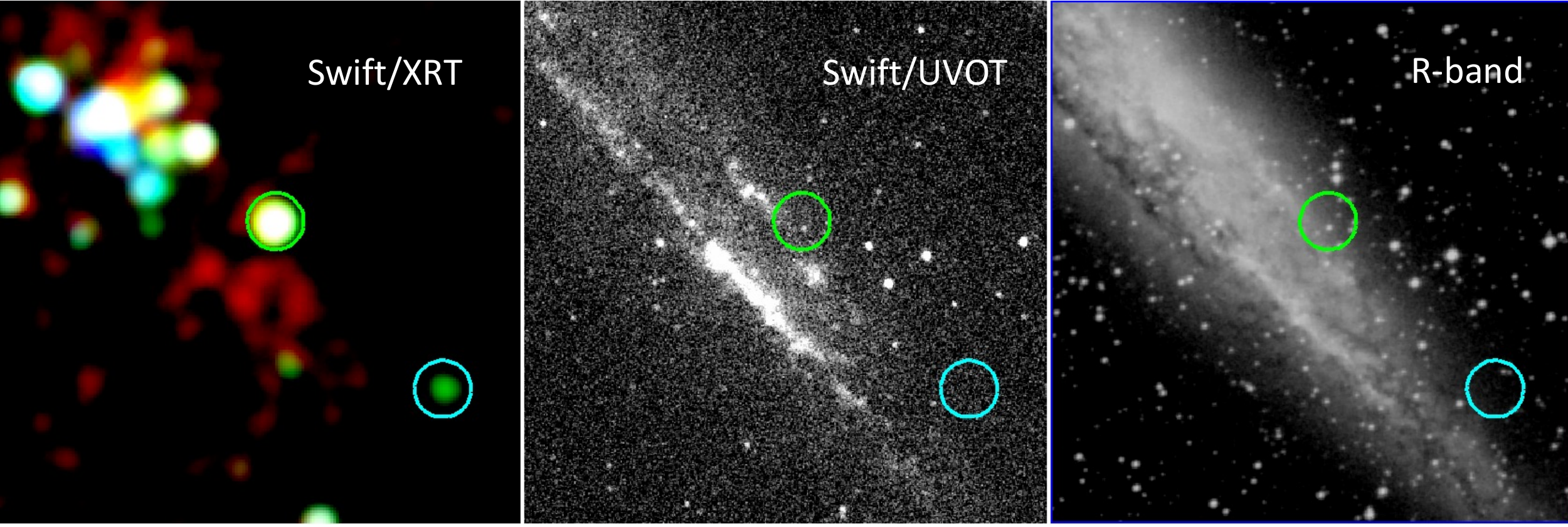}
\end{center}
\caption{\swiftxrt\ (left, red is 0.3--1 keV, green is 1--2.5 keV and blue is 2.5--10 keV, smoothed with a 8\arcsec\ Gaussian), \swiftuvot\ (middle, $UVW2$ filter), and DSS $R$-band image (right) of NGC 4945, with the position of \srci\ marked with a cyan circle and \srcii\ marked with a green circle, both with 25\arcsec\ radius. North is up and East is left.}
\label{fig_J1304_img}
\end{figure*}

\begin{figure}[h]
\begin{center}
\includegraphics[trim=10 20 20 0, width=90mm]{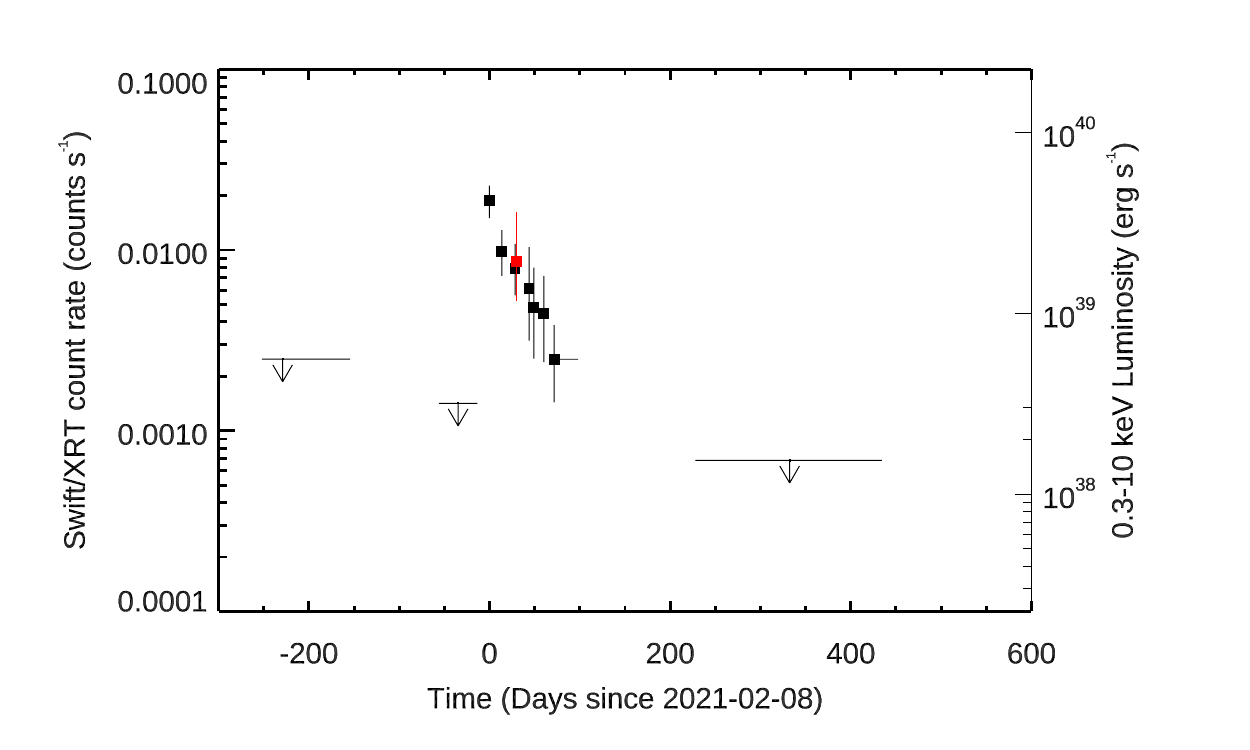}
\end{center}
\caption{\swiftxrt\ lightcurve of \srci, the transient in NGC 4945 (black data points). Upper limits (3$\sigma$) from a stack of observations pre- and post-detection are shown with black arrows. The \chandra\ data are shown in red. The luminosity axis on the right assumes a distance of 3.7 Mpc to the source.}
\label{fig_J1304_ltcrv}
\end{figure}

\srci\ was first detected in a \swiftxrt\ observation taken on 2021 February 8 (obsID 00013908005). The target of the \swift\ observation was NGC 4945 X-1 \citep{brandt96}, an ultraluminous X-ray source hosted by NGC 4945, a barred spiral galaxy in the constellation Centaurus. The enhanced position given by the online tool was R.A. = 196.23411\degree, -49.53306\degree (=13h 04m 56.19s, -49\degree 31\arcmin 59.0\arcsec) with an error radius of 3.2\arcsec (90\% confidence). The position of \srci\ appears to place the source in the outskirts of the galaxy (Figure \ref{fig_J1304_img}). No X-ray source has been reported at this position previously, despite multiple \chandra, \xmm, \suzaku, \nustar\ and \swift\ observations, the last of which was by \swift\ only 2 weeks prior to the new X-ray source being detected, as shown in the lightcurve in Figure \ref{fig_J1304_ltcrv}. After the source was initially detected, it declined in brightness from its peak, becoming undetected by \swiftxrt\ 60 days after its initial detection, even in stacked observations.

We used the online tool to extract the stacked \swiftxrt\ spectrum of the source from 6 observations during which the source was detected. The total exposure time was 12.9 ks. The online tool fitted the spectrum with an absorbed power-law model, which yielded $W=53.02$ with 62 DoFs where \nh$=1.33^{+1.12}_{-0.76}\times10^{22}$ \cmsq\ and $\Gamma=2.63^{+1.06}_{-0.87}$ assuming a Galactic column density of $2.2\times10^{21}$ \cmsq\ \citep{willingale13}. The 0.3-10 keV unabsorbed flux from this model was $1.0^{+4.2}_{-0.6}\times10^{-12}$ \ergcms, which implies a luminosity of $1.7\times10^{39}$ \ergs\ at 3.7 Mpc. The count rate to flux conversion factor was $1.37\times10^{-10}$ erg cm$^{-2}$ count$^{-1}$, which we used to determine the luminosity axis in Figure \ref{fig_J1304_ltcrv}.

The deepest upper limit on the flux of \srci\ prior to its detection is from \chandra\ observations which have a sensitivity of $1.1\times10^{-15}$ \ergcms\ in the 0.5--8 keV band listed in CSC2 \citep{evans10}. This is 3 orders of magnitude lower than the flux measured above. The deepest upper limit from \xmm\ observations is $<7.4\times10^{-15}$ \ergcms\ in the 0.2--12 keV band listed in XSA.

We also obtained a \chandra\ DDT observation of the source which took place on 2021 March 10 (obsID 24986), with ACIS-S at the aimpoint in VFAINT mode. The source was well detected with a count rate of $1.52\times10^{-2}$ \cntrt\ in the 10 ks exposure. We extracted the \chandra\ spectrum with {\sc specextract} from circular regions, radius 1.5\arcsec\ for the source and 7.5\arcsec\ for the background. The spectra were grouped with a minimum of 1 count per bin with the tool {\sc grppha}.

\begin{figure}[h]
\begin{center}
\includegraphics[width=90mm]{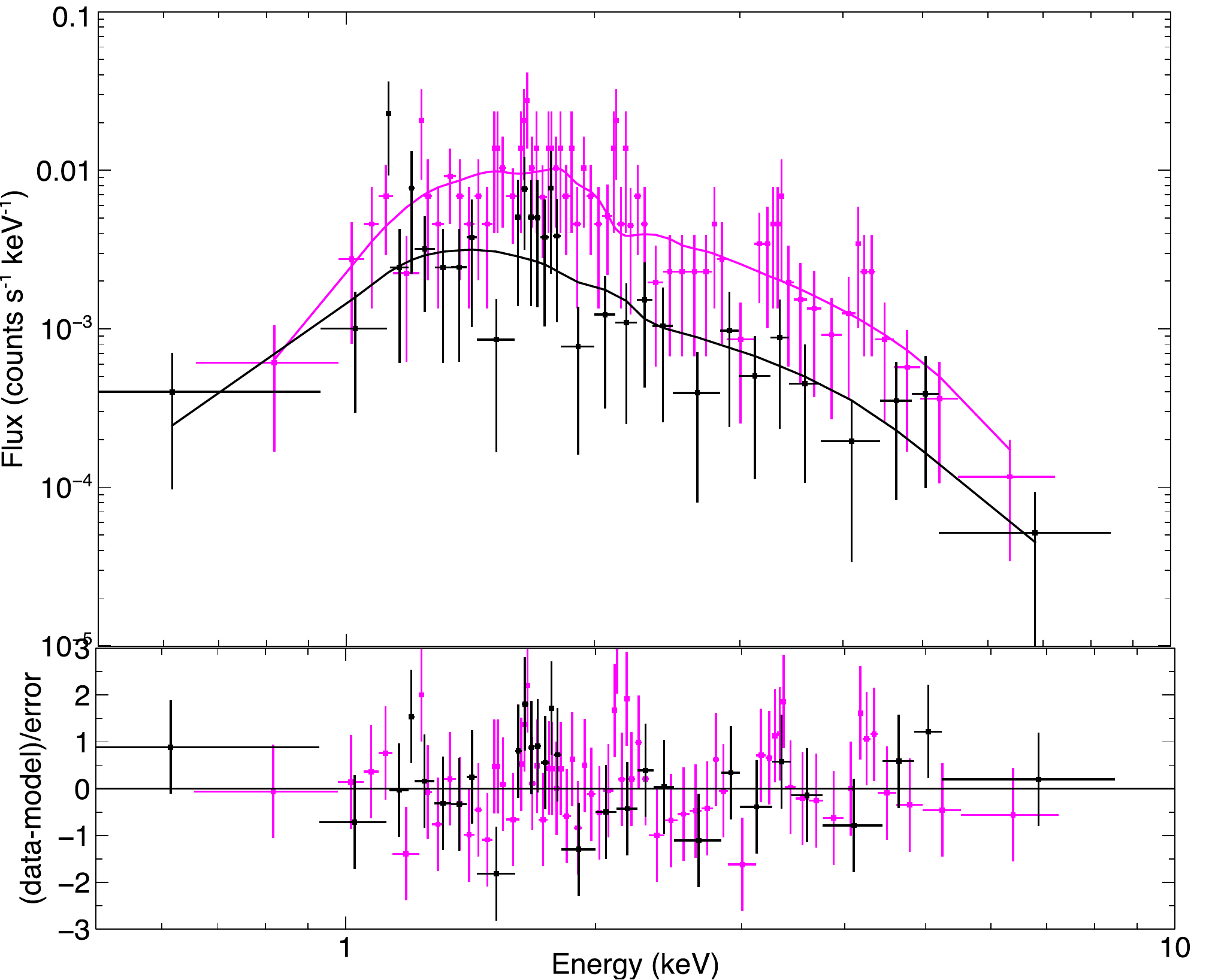}
\end{center}
\caption{\swiftxrt\ (black) and \chandra\ (magenta) spectra of \srci, the X-ray transient in NGC 4945,  fitted simultaneously with an absorbed power-law model with all parameters tied between instruments, but with a cross-normalization constant to allow for differing responses and flux levels.}
\label{fig_J1304_spec}
\end{figure}

We fitted the \chandra\ spectrum of the source with the same model used to fit the \swiftxrt\ spectrum described above. Since we did not find evidence for spectral variability between \swiftxrt\ and \chandra, we fitted the joint \swiftxrt\ and \chandra\ spectrum of the source in {\sc xspec}, with a constant to account for cross-calibration uncertainties and the flux variability of the source. This yielded $W=120.10$ for 174 DoFs. The cross-calibration constant for the \swiftxrt\ spectrum is set to unity, and the constant for the \chandra\ spectrum is $0.70^{+0.21}_{-0.16}$. We find \nh$=1.13^{+0.45}_{-0.38}\times10^{22}$ \cmsq\ and $\Gamma=2.82^{+0.56}_{-0.51}$. The log of the 0.3--10 keV unabsorbed flux from this model corresponding to the time of the \chandra\ observation is $-11.84^{+0.38}_{-0.28}$, which implies a luminosity of $2\times10^{39}$ \ergs\ at 3.7 Mpc. The spectrum is shown in Figure \ref{fig_J1304_spec}.

We also trialled a {\tt diskbb} model in place of the {\tt powerlaw} one, which produced $W=122.14$ for 174 DoFs, a slightly worse fit for the same number of DoFs.  We find \nh$=5.18^{+2.72}_{-2.22}\times10^{21}$ \cmsq\ and $T_{\rm in}=1.03^{+0.24}_{-0.17}$ with a normalization, $N=1.67^{+2.40}_{-1.00}\times10^{-2}$. The normalization is related to the inner disk radius by $R_{\rm in}=D_{10}\times\sqrt{N/cos\theta}$, where $R_{\rm in}$ is the inner disk radius in km, $D_{10}$ is the distance to the source in units of 10 kpc, and $\theta$ is the inclination angle of the disk. Assuming a face-on disk ($\theta=0$) yields $R_{\rm in}=48$ km which is the innermost stable orbit of a 5\msol\ black hole. We note that the luminosity estimate would be a factor of 3.5 lower if this model is assumed and integrated over all energies.

We also used the \chandra\ data to acquire a more precise position of \srci. We compiled an X-ray source list of the \chandra\ observation in the 0.5--8 keV band using {\sc wavdetect} with default parameters and cross-matched this with a {\it Gaia} EDR3 source list of the region \citep{gaia18}, selecting sources within 1.0\arcsec\ of each other, which produced four \chandra/\gaia\ matched sources. We define the astrometric shifts as the mean difference in RA and Dec between these matched sources which is $\delta$RA$=0.34$\arcsec\ and $\delta$Dec$=-0.44$\arcsec. The corrected position is R.A. = 13h 04m 56.350s (196.23479\degree), Decl.=-49\degree\ 31\arcmin\ 59.66\arcsec\ (-49.533239\degree, J2000), which lies in the middle of the \swift\ error circle. The mean residual offset between the corrected \chandra\ positions and the \gaia\ positions is 0.53\arcsec, which we use as our positional error. There are no sources catalogued at other wavelengths within the \chandra\ error circle. The closest source is a near-IR $J=18.9$ source cataloged by the VISTA Hemisphere Survey  \citep[VHS,][]{mcmahon13} and lies 1.68\arcsec\ from the \chandra\ position, and is therefore unlikely to be related. Despite numerous \hst\ observations of NGC 4945, none of them covered the region of the source.

We ran the tool {\tt uvotsource} on the \swiftuvot\ images to obtain photometry of the source in the UV bands using a 2\arcsec\ radius circular region centered on the X-ray position. The source was not detected and we obtained upper limits of $UVW2>22.7$, $UVM2>22.5$ and $UVW1>21.5$ taken from observations when the X-ray source was bright.

\subsection{\srcii, a second X-ray transient in the field of NGC 4945}

\begin{figure}[h]
\begin{center}
\includegraphics[trim=10 20 20 0, width=90mm]{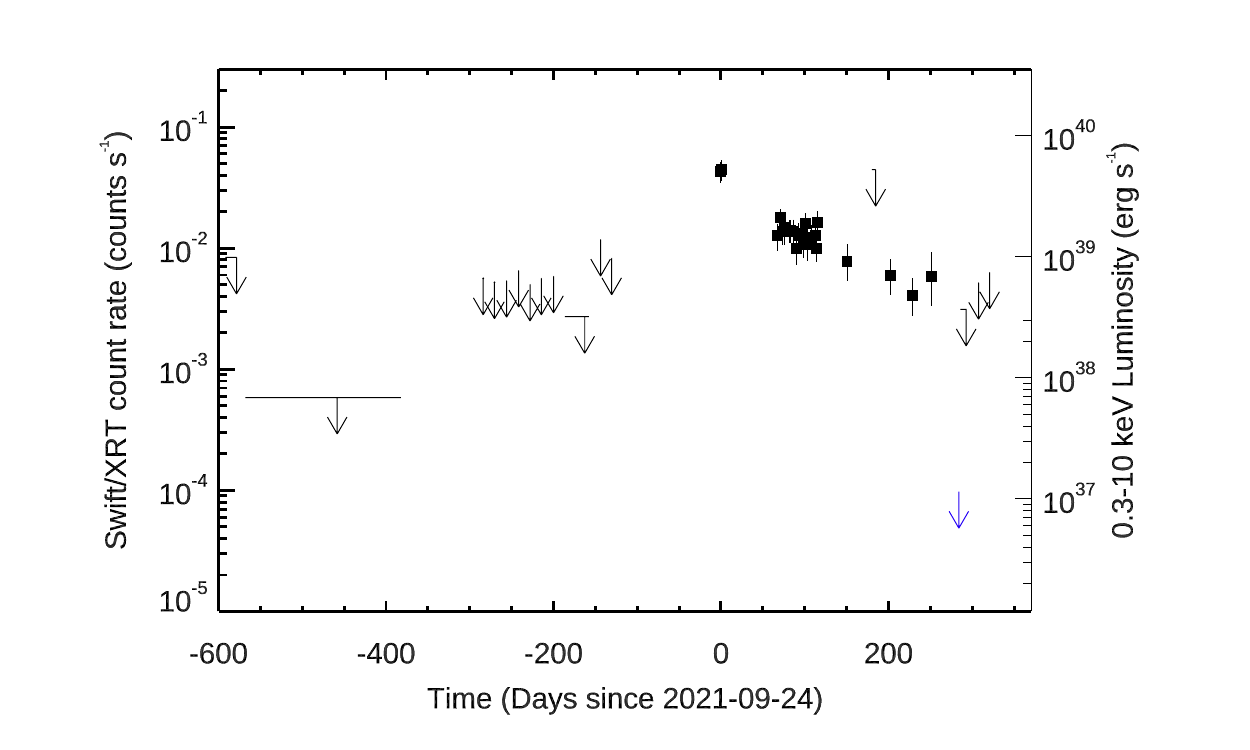}
\end{center}
\caption{\swiftxrt\ lightcurve of \srcii, the second transient in NGC 4945. Upper limits (3$\sigma$) are shown with arrows. Data from \xmm\ are shown in blue. The luminosity axis on the right assumes a distance of 3.7 Mpc to the source.}
\label{fig_J1305_ltcrv}
\end{figure}

This X-ray source was also detected in a \swiftxrt\ observation of NGC 4945 X-1, and was first detected on 2021 September 24 (obsID 00013908017), 7 months after \srci\ as described in Section \ref{sec_J1304} above. The astrometrically corrected position given by the online tool from the first 22 obsIDs where the source was detected was 196.2985\degree, -49.4928\degree\ (=13h 05m 11.65s, -49\degree\ 29\arcmin\ 34.3\arcsec) with an error radius 2.4\arcsec\ (90\% confidence) and we henceforth refer to this source as \srcii. No X-ray source has previously been reported within the positional error circle of \srcii. The \swiftxrt\ lightcurve of the source produced by the online tool is shown in Figure \ref{fig_J1305_ltcrv}, which shows the source declining in brightness until $\sim250$ days after its initial detection, after which the source was undetected by \swiftxrt. The XRT, UVOT and $R$-band images are shown in Figure \ref{fig_J1304_img} which show that similarly to \srci, \srcii\ appears to be in the outskirts of NGC 4945.

We ran the tool {\tt uvotsource} on the \swiftuvot\ images to obtain photometry of the source in the UV and optical bands using a 2\arcsec\ radius circular region centered on the X-ray position. The source was not detected and we obtained upper limits of $UVW2>20.9$, $UVM2>21.2$, $UVW1>20.8$, $U>20.2$, $B>19.5$, and $V>18.8$, taken from obsID 00015017005 taken when the source was X-ray bright.

As with \srci, no source at any wavelength is catalogued within the error region for this X-ray source, and none of the {\it HST} observations of NGC 4945 cover the region. Once again the closest source is a $J=14.6$ mag near-IR source which lies 4.7\arcsec\ from the astrometrically corrected position of the X-ray source, outside the 90\% error circle (2.4\arcsec\ radius).

We used the online tool to extract the stacked \swiftxrt\ spectrum of the source (first 26 observations since detection) with a total exposure time of 45.1 ks. The online tool fitted the spectrum with an absorbed power-law model, which yielded $W=259.57$ with 252 DoFs where \nh$=6.7^{+2.1}_{-1.7}\times10^{21}$ \cmsq\ and $\Gamma=2.23^{+0.30}_{-0.27}$ assuming a Galactic column density of $2.2\times10^{21}$ \cmsq\ \citep{willingale13}. The 0.3-10 keV unabsorbed flux from this model $1.02^{+0.31}_{-0.17}\times10^{-12}$ \ergcms, which implies a luminosity of $1.7\times10^{39}$ \ergs\ at 3.7 Mpc. The count rate to flux conversion factor was $7.71\times10^{-11}$ erg cm$^{-2}$ count$^{-1}$, which we used to determine the luminosity axis in Figure \ref{fig_J1305_ltcrv}.

\begin{figure}[h]
\begin{center}
\includegraphics[width=90mm]{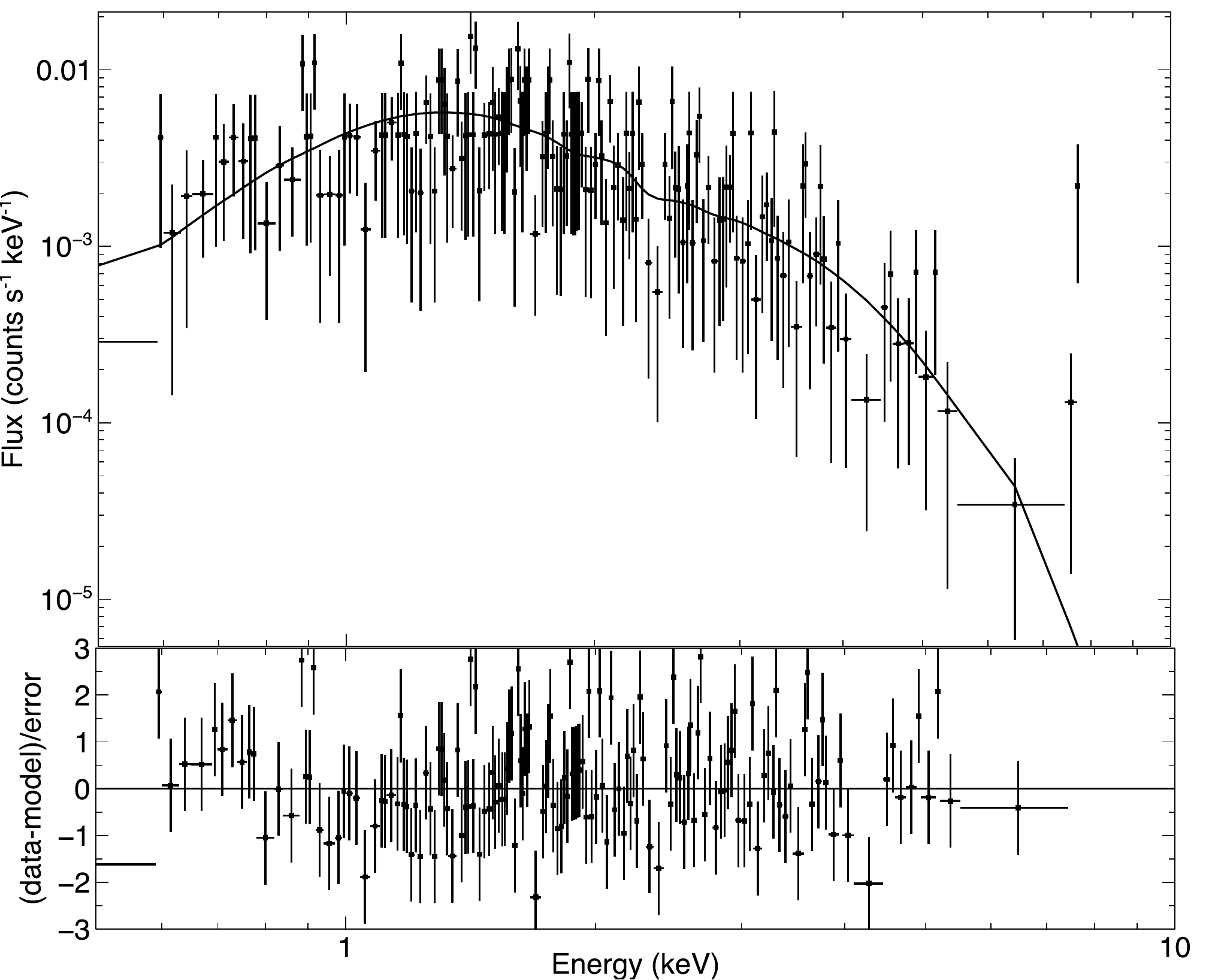}
\end{center}
\caption{\swiftxrt\ spectrum of \srcii, the second X-ray transient in NGC 4945, fitted with an absorbed power-law model.}
\label{fig_J1305_spec}
\end{figure}

Fitting in {\sc xspec}, we found an improvement in the fit could be found with a multicolor disk component ({\tt diskbb}) in the place of the power-law component, which yielded $W$=246.11 with 252 DoFs. The best-fit parameters of this model were \nh$=2.7^{+1.3}_{-1.0}\times10^{21}$ \cmsq, $T_{\rm in}=1.0\pm0.2$ keV, $N=2.3^{+1.7}_{-1.0}\times10^{-2}$.  As for \srci, if we assume a face-on disk we find $R_{\rm in}=56$ km which is the innermost stable orbit of a 6\msol\ black hole. We note that the luminosity estimate would be a factor of 1.9 lower if this model is assumed and integrated over all energies. We plot the spectrum of \srcii\ in Figure \ref{fig_J1305_spec}.

Unfortunately the \chandra\ observation taken of \srci\ as described above did not have \srcii\ in the field of view. The deepest upper limit on the flux of \srcii\ prior to its detection with \swiftxrt\ is from other \chandra\ observations which have a sensitivity of $6.5\times10^{-16}$ \ergcms\ in the 0.5--8 keV band listed in CSC 2.0. This is $>$3 orders of magnitude lower than the flux measured above. The deepest historical upper limit from \xmm\ observations is $<1.1\times10^{-14}$ \ergcms\ in the 0.2--12 keV band listed in XSA.

A 150-ks \xmm\ observation of NGC 4945 took place on 2022 July 5, 284 days after \srcii\ was detected. The XMM data were reduced using a pipeline that utilizes v19.1.0 of the Science Analysis Software (SAS). The {\tt cifbuild} command was used to create a CCF corresponding to the observations, and the {\tt odfingest} command was used to produce a SAS summary file. The data were reduced and MOS and pn event files were created using the {\tt emproc} and {\tt epproc} commands, respectively. We first identify periods of high background by creating a lightcurve of the events in the 10--12 keV band, creating good time intervals where the rate was $<0.4$ \cntrt\ in this band in the pn detector leaving 99\,ks for the pn and 101\,ks for the MOS. Events were selected with $PATTERN\leq4$ for the pn and $PATTERN\leq12$ for the MOS. 

Upon inspection of the images, a faint X-ray enhancement appears at the source location in both the pn and MOS1 data. To test whether this is a detection, spectra were extracted using the {\tt specextract} command with circular source regions of radii 16\arcsec. Local background was accumulated from annuli of the same area just around the source regions. The resulting spectra are heavily background dominated which requires extensive modeling. We therefore calculate an upper limit on the flux of the source by assuming that the spectrum does not change and applying the best-fitting \swiftxrt\ model of a multicolor disk component ({\tt diskbb}) with \nh= $2.7\times10^{21}$ \cmsq, and $T_{\rm in}=1.0$ keV to the source+background spectrum in {\sc xspec}. This yields an upper limit on the 0.3-10 kev flux of $7.5\times10^{-15}$ \ergcms, which implies a 0.3--10 keV unabsorbed luminosity of \lx$=1.2\times10^{37}$, well below the luminosity measured by \swiftxrt\ only 30 days prior (Figure \ref{fig_J1305_ltcrv}).

\subsection{\srciii, an X-ray transient in the field of NGC 7793}
\label{sec_srciii}

\begin{figure*}
\begin{center}
\includegraphics[trim=10 20 20 0, width=180mm]{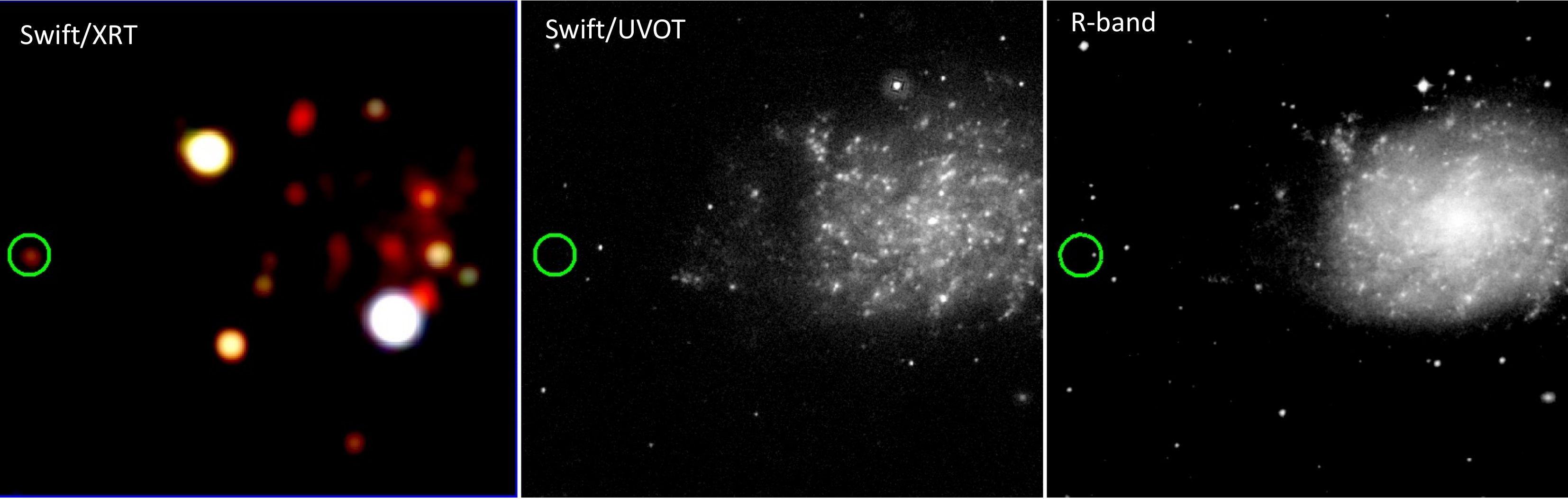}
\end{center}
\caption{\swiftxrt\ (left, red is 0.3--1 keV, green is 1--2.5 keV and blue is 2.5--10 keV, smoothed with a 8\arcsec\ Gaussian), \swiftuvot\ (middle, $U$-band), and DSS $R$-band image (right) of NGC 7793, with the position of \srciii\ marked with a green circle with 25\arcsec\ radius. North is up and East is left.}
\label{fig_ngc7793_img}
\end{figure*}

This X-ray source was first detected in a \swiftxrt\ observation taken on 2018 April 28 (obsID 00094097003) and not found in our real time search, rather a search through archival data. The target of the Swift observation was NGC 7793 P13, an ultraluminous X-ray source hosted by NGC 7793 \citep{read99} known to be a ULX pulsar \citep{fuerst16,israel17}. The enhanced XRT position given by the online tool was R.A.=359.60774\degree, Decl.=-32.60254\degree (=23h 58m 25.86s, -32\degree 36\arcmin 09.2\arcsec) with an error radius of 2.5\arcsec (90\% confidence, Figure \ref{fig_ngc7793_img}). 

The source is listed in the 2SXPS catalogue as \srciii\ with a mean count rate of 5.37$\pm0.76\times10^{-4}$ \cntrt, detected in a stack of data over the date range 2010-08-16--2018-07-28. This average count rate was 2 orders of magnitude below the newly detected count rate. We note that this average flux is from a date range that covers periods both when the source was undetected in individual observations and when it was detected in individual observations.

\begin{figure}[h]
\begin{center}
\includegraphics[trim=10 20 20 0, width=90mm]{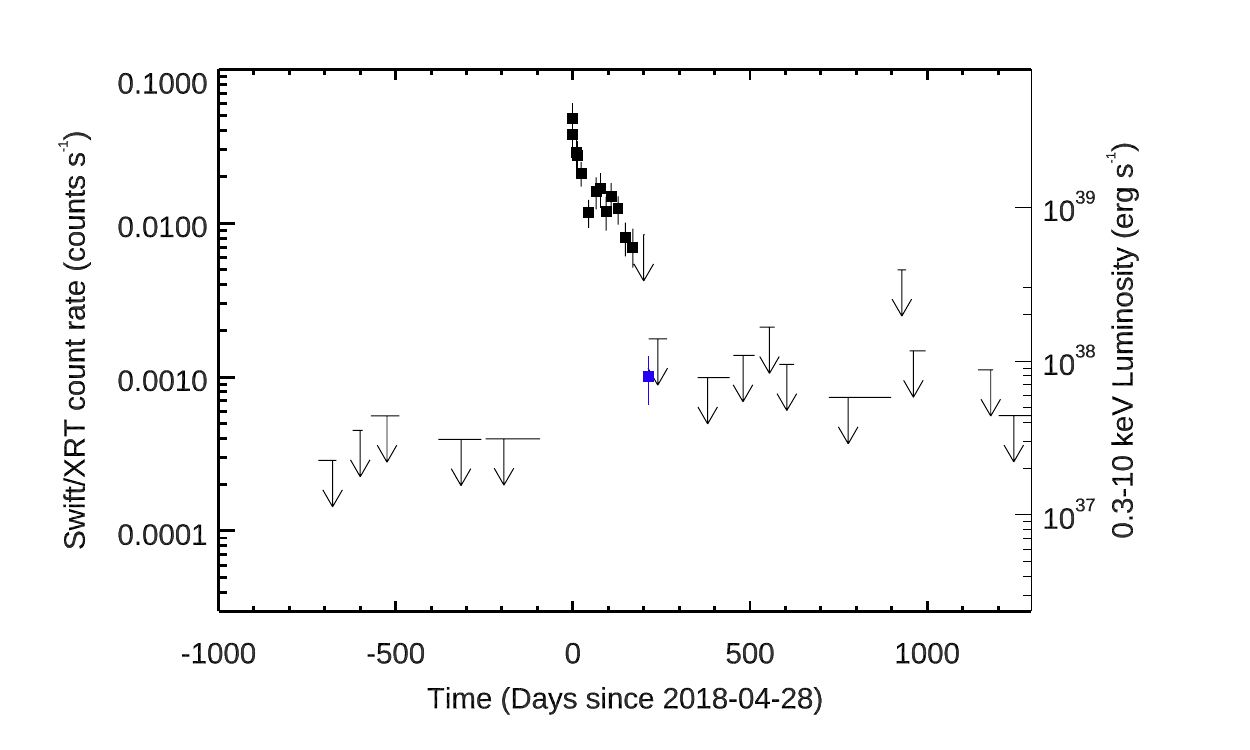}
\end{center}
\caption{\swiftxrt\ lightcurve of \srciii, the transient in the field of NGC 7793. Upper limits (3$\sigma$) are shown with downward pointing arrows. The luminosity axis on the right assumes a distance of 3.8 Mpc to the source. Data from \xmm\ are shown in blue.}
\label{fig_ngc7793_ltcrv}
\end{figure}

Since this source was first catalogued in 2SXPS, we henceforth refer to it by its catalogued name \srciii. The lightcurve produced by the online tool is shown in Figure \ref{fig_ngc7793_ltcrv}. This shows that prior to 2018 April 28, the source was not detected in stacked observations with upper limits consistent with the 2SXPS count rate. The source was not detected in the XRT observation immediately preceding April 28 on April 22. After reaching its peak, the source declined monotonically until it was no longer detected by \swiftxrt\ 180 days afterwards. 

 A {\tt diskbb} model in place of the {\tt powerlaw} one produced an improvement in the fit of $\Delta$C=-9, where \nh$<5.9\times10^{20}$ \cmsq\ and $T_{\rm in}=1.03^{+0.21}_{-0.16}$ with a normalization, $N=1.72^{+1.65}_{-0.87}\times10^{-2}$. The normalization corresponds to $R_{\rm in}=50$ km which is the innermost stable orbit of a 6\msol\ black hole when assuming a face on disk. We note that the luminosity estimate would be a factor of 1.7 lower if this model is assumed and integrated over all energies.

An \xmm\ observation took place on 2018-11-27, 213 days after the initial detection by \swiftxrt\ (obsID 0823410301). We filter the data in the same way as described for \srcii, which results in 17.6 ks of data. A circular region with a radius of 15\arcsec\ was used to extract the source spectrum, and an annulus with inner radius 25\arcsec\ and outer radius of 45\arcsec\ was used to extract the background spectrum. The data were grouped with a minimum of 1 count per bin using {\sc grppha}. The source was background dominated above 1 keV so we excluded these channels and the resulting average count rate in the 0.2--1 keV band was 2.5$\pm0.5\times10^{-3}$ \cntrt.  

\begin{figure}[h]
\begin{center}
\includegraphics[width=90mm]{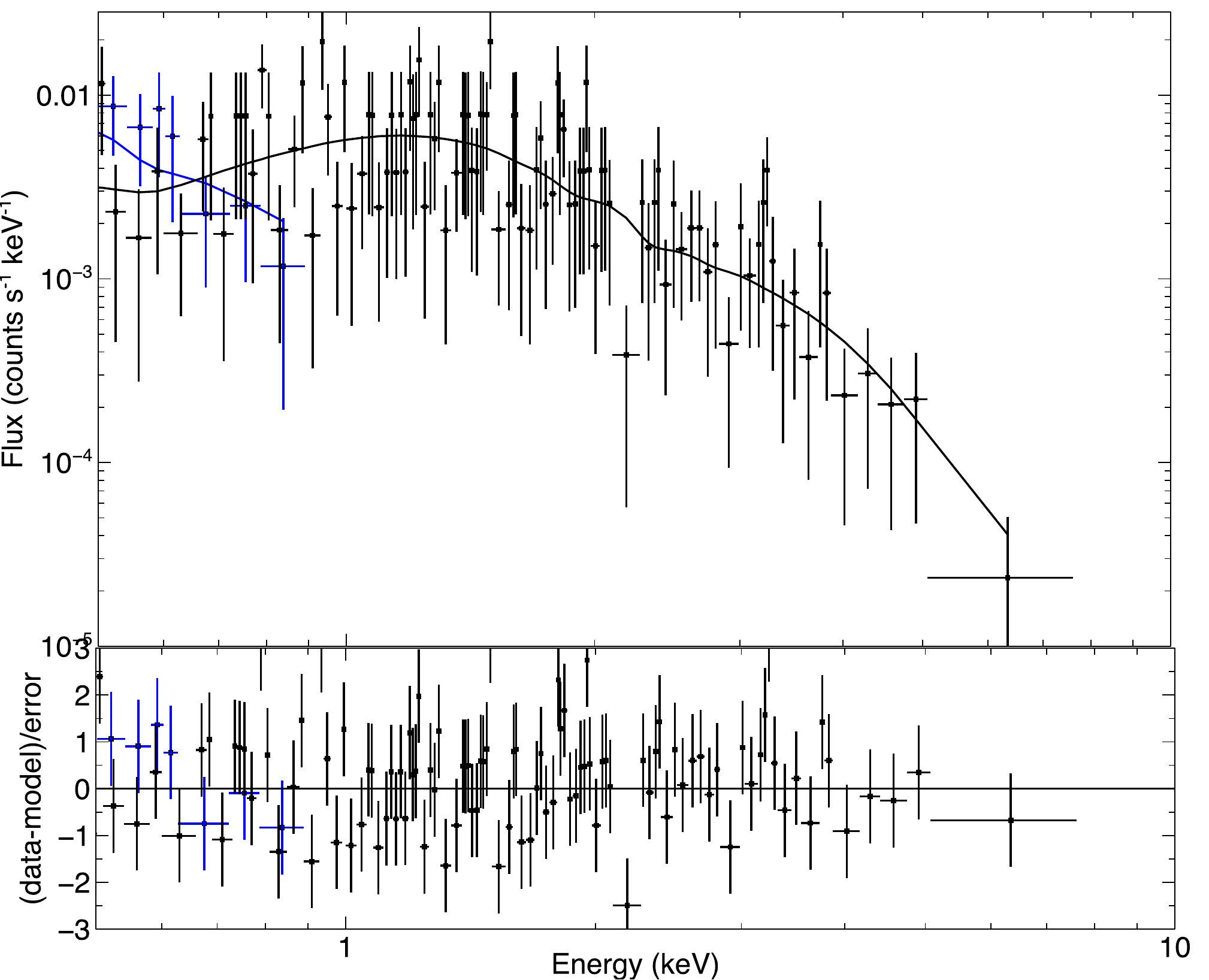}
\end{center}
\caption{\swiftxrt\ (black) and \xmm\ (blue) spectra of \srciii, the X-ray transient in NGC 7793,  fitted simultaneously with an absorbed power-law model with all parameters tied between instruments, but with a cross-normalization constant to allow for differing responses and flux levels.}
\label{fig_2SXPS_4274_spec}
\end{figure}

Due to the narrow bandpass, we fit the \xmm\ spectrum with the same model as for the \swift\ spectrum, with all parameters fixed with the exception of the normalization, which yielded $N=1.4\times10^{-5}$ with $W=73.04$ with 54 DoFs. The 0.3--10 keV flux is 4.6$^{+1.6}_{-1.5}\times10^{-14}$ \ergcms, which implies a luminosity of 7.9$^{+1.8}_{-2.6}\times10^{37}$ \ergs\ at 3.8 Mpc \citep{sabbi18}. We plot this flux in Figure \ref{fig_ngc7793_ltcrv}. The source was not detected in an observation only 1 month after the above \xmm\ observation (obsID 0823410401 on 2018-12-27). The upper limit on the 0.2--10 keV flux is listed as 5.9$\times10^{-15}$ \ergcms\ in 4XMM from this observation, with similar upper limits provided by further observations since then. We show the \swift\ and \xmm\ spectra in Figure \ref{fig_2SXPS_4274_spec}.

The deepest upper limit from \xmm\ observations prior to the \swift\ detection is $<5.0\times10^{-15}$ \ergcms\ in the 0.2--12 keV band listed in XSA. This is 2 orders of magnitude lower than the peak flux measured above.

4XMM DR11 lists the source from obsID 0823410301 as 4XMM J235825.9-323610 at R.A.=23h 58m 25.98s and Decl.=-32\degree 36\arcmin 10.5\arcsec\ with a positional error of 1.0\arcsec, which is an improvement on the XRT position. 

Fortuitously, \hst\ has observed the region of \srciii\ as part of the GHOSTS survey \citep{radburnsmith11} with the ACS and F606W and F814W filters. However, no source is listed in the {\it Hubble Source Catalogue} (v3) within the \xmm\ positional error and the closest source lies 2.1\arcsec\ away.

We ran {\sc uvotsource} on the mostly $U$-band UVOT data but did not detect the source in any observation to a limiting magnitude of $U\sim21.5$.

\subsection{\srcv, a second X-ray transient in NGC 7793}

\begin{figure*}
\begin{center}
\includegraphics[trim=10 20 20 0, width=120mm]{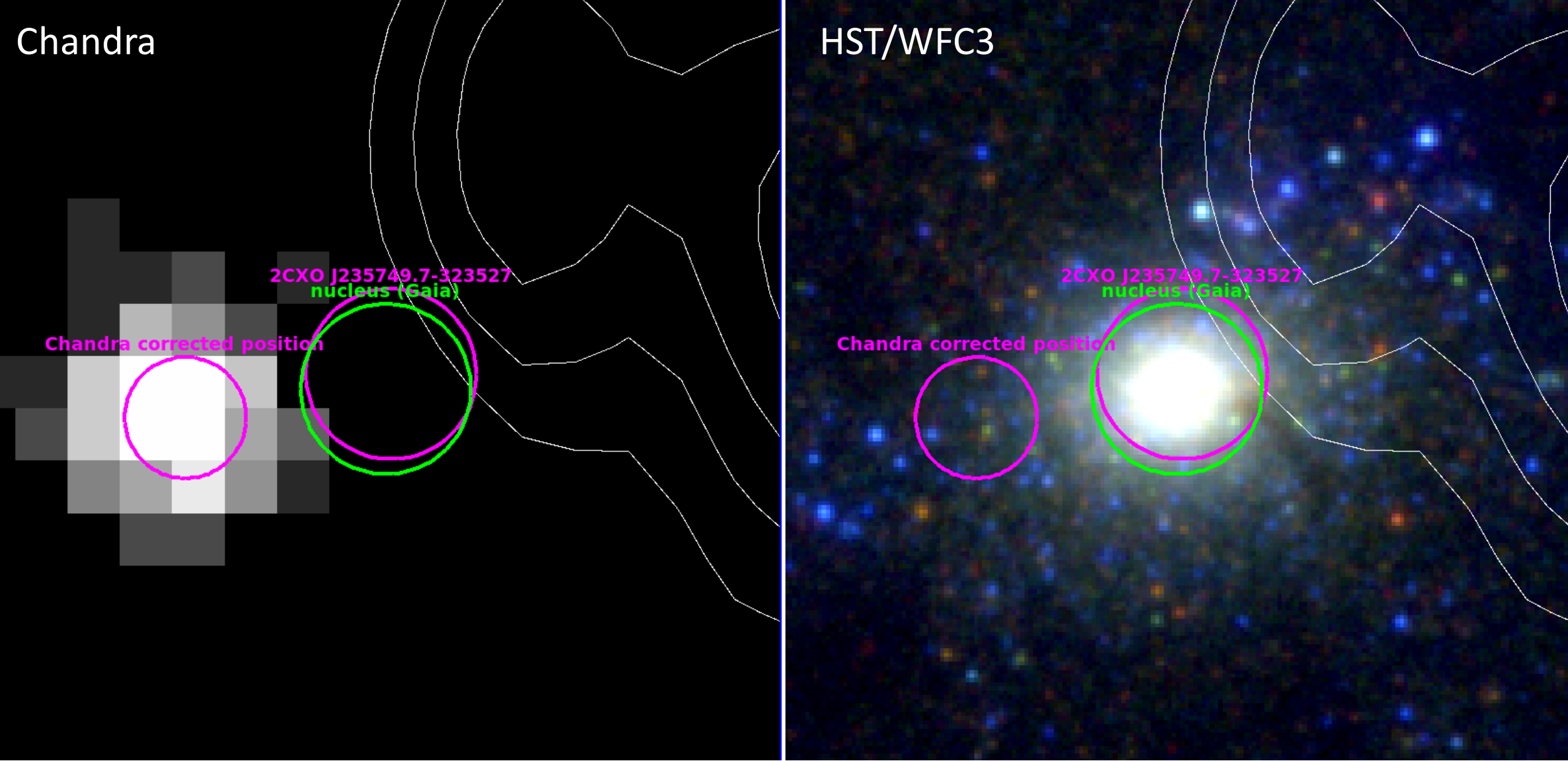}
\end{center}
\caption{ \chandra\ (left) and \hst/WFC3/UVIS (right, red is $F814W$, green is $F547M$ and blue is $F275W$) image of the nuclear region of NGC 7793, with the corrected \chandra\ position of \srcv, the X-ray transient, marked with a magenta circle. The \gaia\ position of the nucleus is shown with a 0.8\arcsec\ green circle. 2CXO J235749.7-323527, a previously catalogued X-ray source coincident with the nucleus is also marked with a 0.8\arcsec\ magenta circle. The white contours show the VLA radio emission (-3, 3 4, 5 ,6 and 7$\times$rms, where rms = 1$\times10^{-5}$ Jy/beam). North is up and East is left.}
\label{fig_srcv_img}
\end{figure*}

\begin{figure}[h]
\begin{center}
\includegraphics[trim=10 20 20 0, width=90mm]{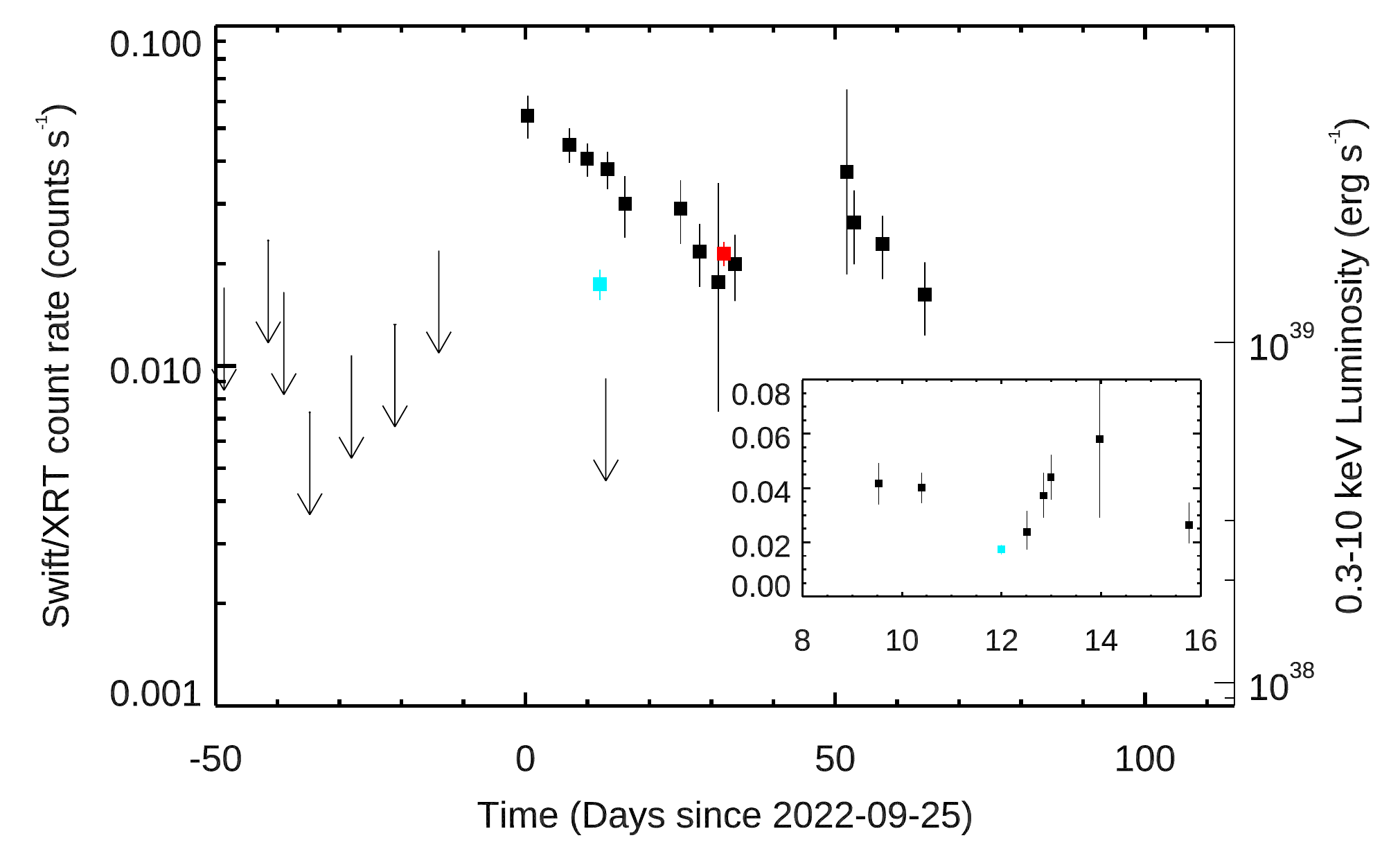}
\end{center}
\caption{\swiftxrt\ lightcurve of \srcv, the second transient in NGC 7793. Upper limits (3$\sigma$) are shown with black arrows. \nustar\ and \chandra\ data are shown in cyan and red respectively. The inset shows a zoom in around the time of the \nustar\ observation with the \swift\ data binned by snapshot to show that the source showed short term variability during this time. The luminosity axis on the right assumes a distance of 3.7 Mpc to the source.}
\label{fig_srcv_ltcrv}
\end{figure}

This X-ray source was also detected in a \swiftxrt\ observation of NGC 7793 P13, and was first detected on 2022 September 25 (obsID 00031791173), 4 years and 5 months after \srcii\ (see Section \ref{sec_srciii} above). The enhanced position given by the online tool from the first 9 obsIDs where the source was detected was 359.458\degree, -32.590806\degree\ (=23h 57m 49.92s, -32\degree\ 35\arcmin\ 26.9\arcsec) with an error radius 2.5\arcsec\ (90\% confidence) and we henceforth refer to this source as \srcv\ (Figure \ref{fig_srcv_img}). A faint \chandra\ source, 2CXO J235749.7-323527, has previously been reported within the positional error circle of \srcv\ with a flux of $2.2\times10^{-15}$ \ergcms\ in the 0.5--7 keV band, 3 orders of magnitude lower than the inferred XRT flux, which is also coincident with the {\it Gaia} nuclear position of the galaxy (Figure \ref{fig_srcv_img}). The \swiftxrt\ lightcurve of the source produced by the online tool is shown in Figure \ref{fig_srcv_ltcrv}, which shows the source declining in brightness from its initial detection. A lightcurve binned by snapshot rather than observation is also shown to highlight some short-term variability seen.

We used the online tool to extract the stacked \swiftxrt\ spectrum of the source (first 9 observations since detection) with a total exposure time of 14 ks. The online tool fitted the spectrum with an absorbed power-law model, which yielded $W=176.66$ with 204 DoFs where \nh$=2.1^{+0.9}_{-0.8}\times10^{21}$ \cmsq\ and $\Gamma=2.02^{+0.25}_{-0.24}$ assuming a Galactic column density of $1.2\times10^{20}$ \cmsq\ \citep{willingale13}. The 0.3-10 keV unabsorbed flux from this model $1.72^{+0.26}_{-0.20}\times10^{-12}$ \ergcms, which implies a luminosity of $3.0\times10^{39}$ \ergs\ at 3.8 Mpc. The count rate to flux conversion factor was $4.86\times10^{-11}$ erg cm$^{-2}$ count$^{-1}$, which we used to determine the luminosity axis in Figure \ref{fig_srcv_ltcrv}.

We obtained a \nustar\ DDT observation of \srcv\ which occurred on 2022 October 7 (obsID 90801526002) with an exposure of 53 ks. We used {\sc heasoft} v6.28, {\sc nustardas} v2.0.0 and {\sc caldb} v20211115 to analyze the data. We produced cleaned and calibrated events files using {\sc nupipeline} with the default settings on mode 1 data only. We used {\sc nuproducts} to produce spectral data, including source and background spectra, and response files. A circular region with a radius of 40\arcsec\ was used to extract the source spectra and a radius of 80\arcsec\ was used to extract the background spectra, taking care to extract the background from the same chip as the source. The source is detected with a count rate of $5\times10^{-3}$ \cntrt\ in the 3--10 keV band in each FPM, above which the background dominates the source. We used the absorbed power-law model described above to determine the flux plotted in Figure \ref{fig_srcv_ltcrv}.

We also obtained a \chandra\ DDT observation of the source which took place on 2022 October 27 (obsID 27481), with ACIS-S at the aimpoint in VFAINT mode. The source was well detected with a count rate of $4.09\times10^{-2}$ \cntrt\ in the 10 ks exposure. We extracted the \chandra\ spectrum with {\sc specextract} from a circular region, radius 2.0\arcsec\ for the source and an annulus radii of 3.1 and 6.2\arcsec\ for the background. The spectra were grouped with a minimum of 1 count per bin with the tool {\sc grppha}. Again we used the absorbed power-law model from the \swift\ data to determine the flux plotted in Figure \ref{fig_srcv_ltcrv}.

\begin{figure}[h]
\begin{center}
\includegraphics[width=90mm]{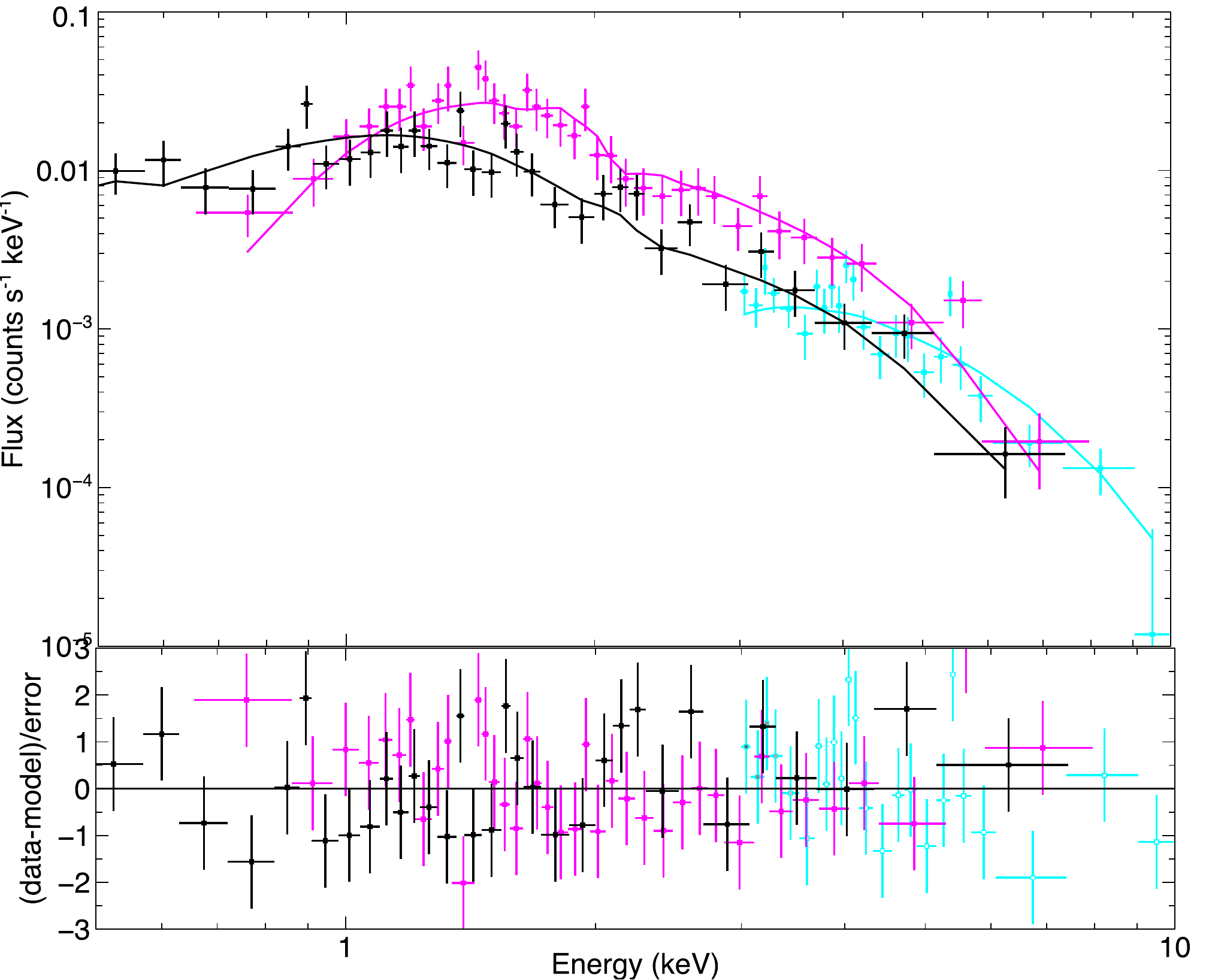}
\end{center}
\caption{\swiftxrt\ (black), \nustar\ (cyan, FPMA and FPMB combined for plotting purposes) and \chandra\ (magenta) spectra of \srcv, the second X-ray transient in NGC 7793. These have been fitted simultaneously with an absorbed multicolor disk black body model with all parameters tied between instruments, but with a cross-normalization constant to allow for differing responses and flux levels.}
\label{fig_srcv_spec}
\end{figure}

We then fitted the \swift, \nustar\ and \chandra\ data simultaneously initially with an absorbed power-law model in {\sc xspec}, with a constant to account for cross-calibration uncertainties and the flux variability of the source. This yielded $W=580.10$ for 611 DoFs. However, this revealed structure in the data to model residuals indicating that a more complex model was required. We then trialled the addition of a high energy cut off to the power-law model ({\tt cutoffpl}) which lead to an improved $W=534.30$ for 610 DoFs. A multicolor disk blackbody model, {\tt diskpbb}, similarly produced $W=534.00$ for 610 DoFs, which we select as our best-fit model due to the slightly better fit statistic. The best-fit parameters are intrinsic line-of-sight column density \nh$=1.2^{+0.7}_{-0.8}\times10^{21}$ \cmsq, inner disk temperature $T_{\rm in}=1.24^{+0.17}_{-0.16}$ keV, disk temperature index $p=0.54^{+0.10}_{-l}$ (where $_{-l}$ indicates the lower bound uncertainty reached the lower limit of 0.5 for the parameter) and normalization $N=8^{+15}_{-5}\times10^{-3}$.  the total luminosity of this model is $3.7\times10^{39}$ \ergs. Given the normalization of the {\tt diskpbb} model and assuming a face-on inclination of the disk ($\theta=0$\degree), the implied black hole mass is 4~\msol\ for a non-spinning black hole. The spectra fitted with this model are shown in Figure \ref{fig_srcv_spec}. We also checked for variation of the parameters between observations by untying them in the fit, but did not find any evidence for this.

We also used the \chandra\ data to acquire a more precise position for \srcv\ in the same way as was done for \srci, which produced seven \chandra/\gaia\ matched sources. The astrometric shifts were $\delta$RA$=+0.34$\arcsec\ and $\delta$Dec$=+0.22$\arcsec. After subtracting these shifts, the corrected position is R.A. = 23h 57m 49.903s (359.45793\degree), Decl.=-32\degree\ 35\arcmin\ 27.97\arcsec\ (-32.591104\degree, J2000), which lies within the \swift\ error circle. The mean residual offset between the corrected \chandra\ positions and the \gaia\ positions is 0.57\arcsec, which we use as our positional error. With this improved positional uncertainty, 2CXO J235749.7-323527 and the nucleus of NGC 7793 are excluded as counterparts to this new X-ray source since they lie 2.0\arcsec\ away (Figure \ref{fig_srcv_img}).  2CXO J235749.7-323527 was not detected in this observation, however. Its reported flux was around the limiting flux of the new observation, which is the likely reason for the non-detection. There are also no other sources catalogued at other wavelengths within the \chandra\ error circle for \srcv, with the exception of eight Hubble Source Catalog (HSC) v3 sources \citep{whitmore16} which we will discuss in Section \ref{sec_hst}.

Finally, due to the possibility that \srcv\ was a nuclear transient, we obtained radio follow-up of the source with the Very Large Array (VLA). The VLA observations were carried out on 2022 October 27 at X-band (8-12 GHz), in C configuration. The angular resolution was 5.9\arcsec$\times2.3$\arcsec, slightly larger than the nominal one due to the low declination of the source. The field of view included the entire host galaxy structure. We did not detect radio emission at the position of the source obtained by \chandra, resulting in a 3-$\sigma$ upper limit of 18 $\mu$Jy/beam. An emitting region is visible starting $\sim$3\arcsec\ towards west from the transient position ( Fig \ref{fig_srcv_img}), with an angular size of about 13\arcsec\ corresponding with optical emission from the nuclear star cluster \citep[e.g.][]{carson15,mondal21}.

\subsection{\srciv, an X-ray transient in the field of M81}
\begin{figure*}
\begin{center}
\includegraphics[trim=10 20 20 0, width=180mm]{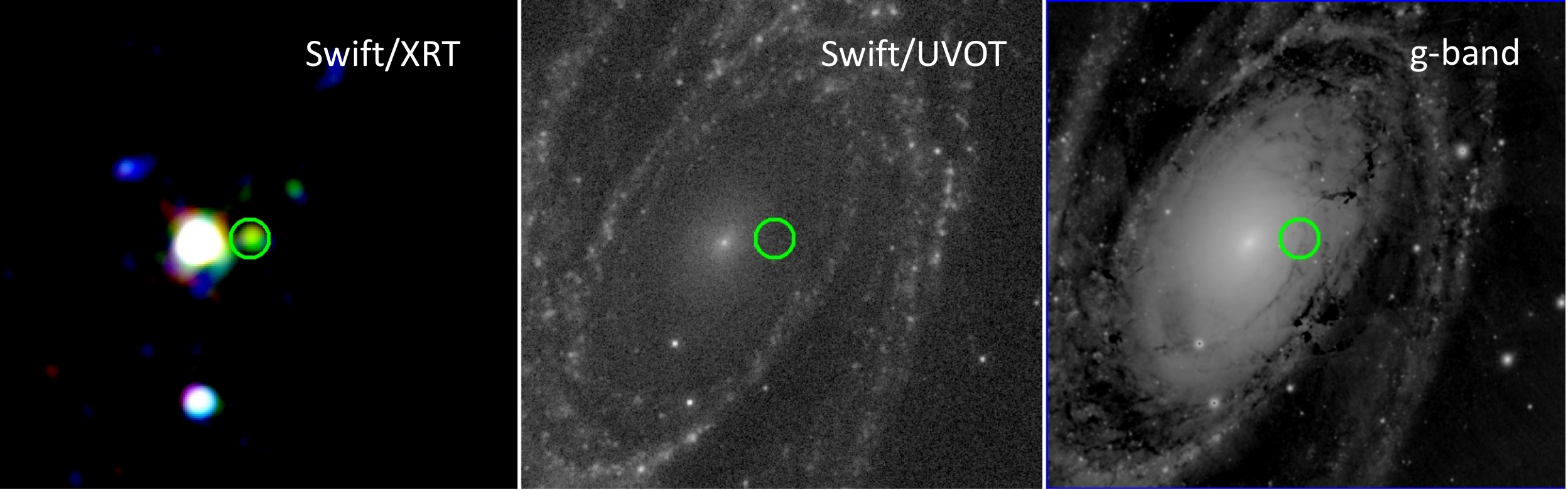}
\end{center}
\caption{\swiftxrt\ (left, red is 0.3--1 keV, green is 1--2.5 keV and blue is 2.5--10 keV, smoothed with a 8\arcsec\ Gaussian), \swiftuvot\ (middle, $UVW2$-band), and PanSTARRS $g$-band image (right) of M81, with the position of \srciii\ marked with a green circle with 25\arcsec\ radius. North is up and East is left.}
\label{fig_M81_img}
\end{figure*}

This X-ray source was first detected in a \swiftxrt\ observation taken on 2022 April 3 (obsID 00096886002). The target of the Swift observation was M81, a Seyfert 2 galaxy. The enhanced position given by the online tool was 148.83625\degree, 69.06692\degree\ (=09h 55m 20.70s, +69\degree\ 04\arcmin\ 00.9\arcsec), with an error radius of 3.2\arcsec (90\% confidence) which appeared to place the source within the galaxy 1.1\arcmin\ from the nucleus (Figure \ref{fig_M81_img}). We will henceforth refer to this source as \srciv. No X-ray source had been reported at this position previously, despite multiple \chandra, \xmm, \nustar\ and \swift\ observations, the last of which was by \swift\ 2 days prior to the new X-ray source being detected, albeit in a short ($<$200\,s) observation. 

Since this source is close to the bright nucleus of M81, we do not use the automated online tool to generate the spectrum and lightcurve as for the other sources. This is to ensure that the nucleus is properly accounted for. We therefore download the observations and extracted events of the source using the {\sc heasoft} v6.25 tool {\sc xselect} \citep{arnaud96}. Source events were selected from a circular region with a 25\arcsec\ radius centered on the above coordinates. Background events were also selected from a circular region with a 25\arcsec\ radius placed at the same distance from the nucleus as the source in order to sample the PSF at its position. For each source spectrum, we constructed the auxiliary response file (ARF) using {\tt xrtmkarf}. The relevant response matrix file (RMF) from the CALDB was used. All spectra were grouped with a minimum of 1 count per bin for spectral fitting purposes.

\begin{figure}[h]
\begin{center}
\includegraphics[trim=10 20 20 0, width=90mm]{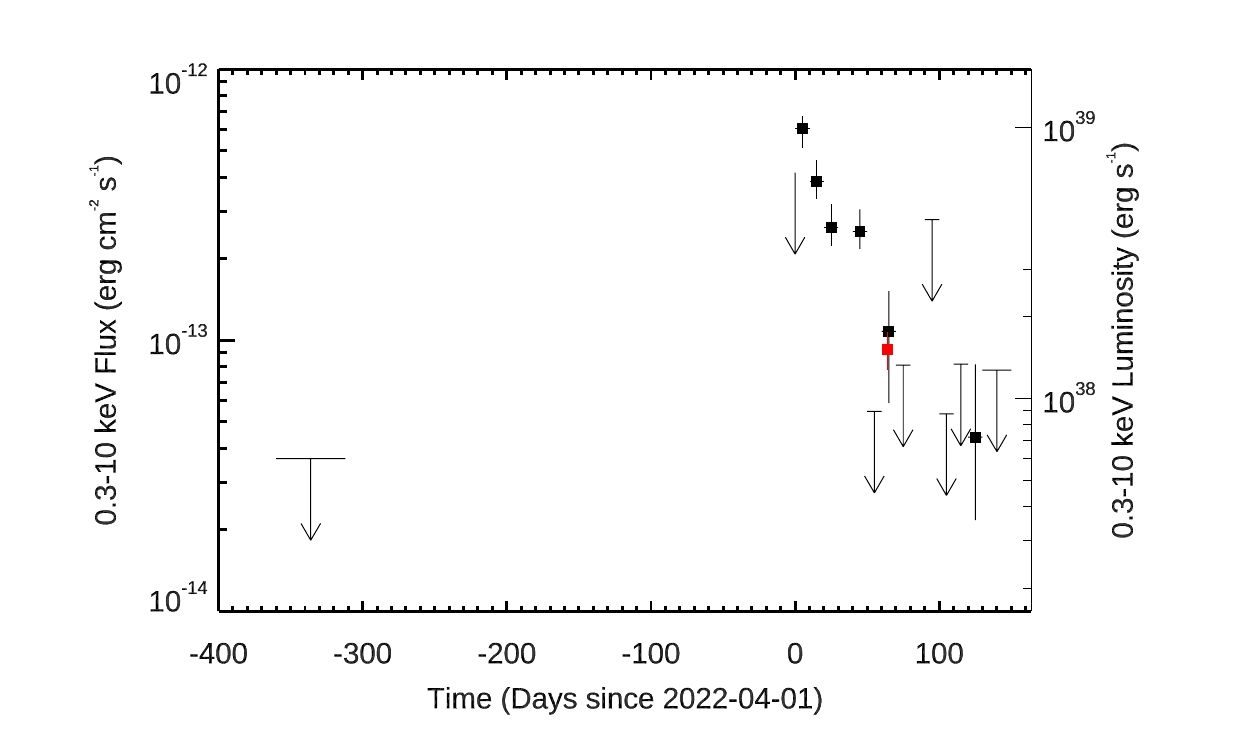}
\end{center}
\caption{\swiftxrt\ lightcurve of \srciv, the transient in M81 (black data points), with the \chandra\ flux data point in red. The luminosity axis on the right assumes a distance of 3.7 Mpc to the source.}
\label{fig_M81_ltcrv}
\end{figure}

We start by simultaneously fitting the \swiftxrt\ spectra from the first 15 observations where the source was detected. We fitted the spectra with an absorbed power-law model with a constant applied to account for the variability of the source from observation to observation. This yielded $W=78.24$ with 94 DoFs where \nh$<5.2\times10^{21}$ \cmsq\ and $\Gamma=2.0^{+1.9}_{-0.6}$. To produce the lightcurve, we stack the individual observations in time bins of 10 days using the tool {\sc addascaspec} and again group the stacked spectra with a minimum of 1 count per bin. We then fit these with the above model where \nh\ and $\Gamma$ are frozen. We plot the flux and implied luminosity in Figure \ref{fig_M81_ltcrv}.

\begin{figure}[h]
\begin{center}
\includegraphics[width=90mm]{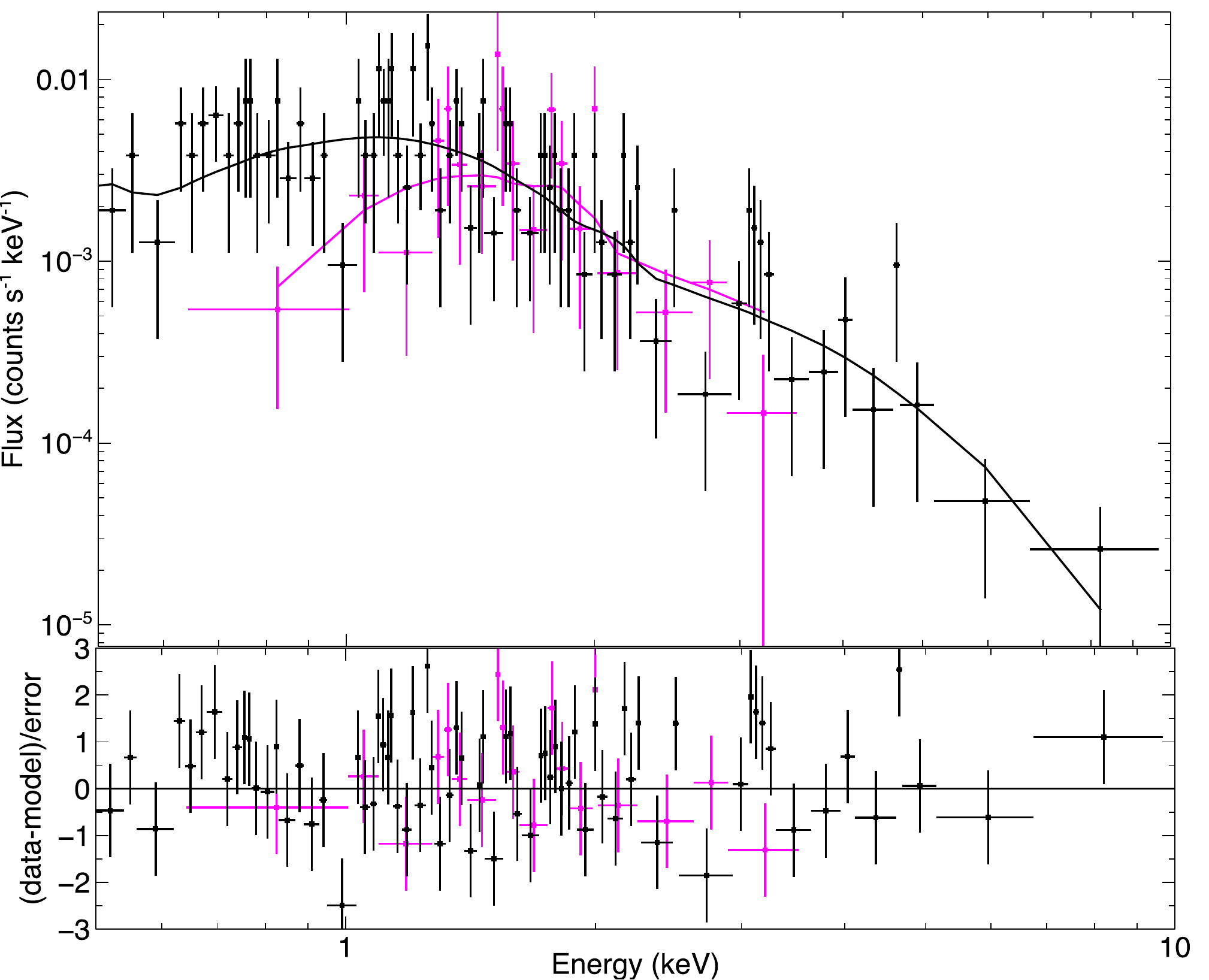}
\end{center}
\caption{\swiftxrt\ (black) and \chandra\ (magenta) spectra of \srciv, the X-ray transient in M81,  fitted simultaneously with an absorbed power-law model with all parameters tied between instruments, but with a cross-normalization constant to allow for differing responses and flux levels.}
\label{fig_M81_spec}
\end{figure}

We also obtained a \chandra\ DDT observation of the source which took place on 2022 June 04 (obsID 24621), with ACIS-S at the aimpoint in FAINT mode. The source was well detected with a count rate of $3.89\times10^{-2}$ \cntrt\ in the 10 ks exposure. We extracted the \chandra\ spectrum with {\sc specextract} from a circular region, radius 2.0\arcsec\ for the source and an annulus radii of 2.5 and 12 \arcsec\ for the background. The spectra were grouped with a minimum of 1 count per bin with the tool {\sc grppha}. We fitted the spectrum with an absorbed power-law model as done for the \swiftxrt\ data which yielded $W=27.66$ with 35 DoFs where \nh$<2.8\times10^{22}$ \cmsq\ and $\Gamma=4.6^{+3.5}_{-2.6}$, consistent with \swiftxrt, albeit with large uncertainties. If we fit the stacked \swiftxrt\ data and the \chandra\ data together we find $W=109.49$ with 128 DoFs where \nh$<6.8\times10^{21}$ \cmsq\ and $\Gamma=2.6^{+1.6}_{-1.1}$. We show these spectra in Figure \ref{fig_M81_spec}.

 A {\tt diskbb} model in place of the {\tt powerlaw} one produced a worsened fit with $\Delta$C=10, where \nh$<5\times10^{20}$ \cmsq\ and $T_{\rm in}=1.03^{+0.18}_{-0.16}$ with a normalization, $N=1.17^{+1.20}_{-0.52}\times10^{-2}$. The normalization corresponds to $R_{\rm in}=40$ km which is the innermost stable orbit of a 5 \msol\ black hole when assuming a face on disk. We note that the luminosity estimate would be a factor of 1.8 lower if this model is assumed and integrated over all energies.

The deepest upper limit on the flux of \srciv\ prior to its detection with \swiftxrt\ is from \chandra\ observations which have a sensitivity of $9.8\times10^{-16}$ \ergcms\ in the 0.5--8 keV band listed in CSC 2.0. This is 3 orders of magnitude lower than the flux measured above. The deepest upper limit from \xmm\ observations is $<1.7\times10^{-12}$ \ergcms\ in the 0.2--12 keV band listed in XSA and is from a slew.

We also used the \chandra\ data to acquire a more precise position for \srciv\ in the same way as was done for \srci\ and \srcv, which produced nine \chandra/\gaia\ matched sources. The astrometric shifts were $\delta$RA$=-0.85$\arcsec\ and $\delta$Dec$=-0.75$\arcsec. After subtracting these shifts, the corrected position is R.A. = 09h 55m 20.873s (148.83697\degree), Decl.=+69\degree\ 04\arcmin\ 02.53\arcsec\ (+69.06737\degree, J2000), which lies towards the edge of the \swift\ error circle. The mean residual offset between the corrected \chandra\ positions and the \gaia\ positions is 0.33\arcsec, which we use as our positional error. There are no sources catalogued at other wavelengths within the \chandra\ error circle, with the exception of 3 Hubble Source Catalog (HSC) v3 sources \citep{whitmore16} which we will discuss in Section \ref{sec_hst}.

\section{What is the nature of these sources?}

\subsection{Foreground sources in our Galaxy?}

If these transient X-ray sources are within our Galaxy, the implied peak luminosities would be $\sim10^{33}$ \ergs\ if assuming a distance of 10 kpc. On the timescale-luminosity plot of \cite{polzin22} the source types most consistent with this luminosity and timescales of $10^2$ days are classical/dwarf novae, however novae are usually accompanied by optical/UV emission \citep[e.g.][]{page20}. This luminosity is comparable to X-ray flares from stars, but the X-ray activity lasted much longer than typical stellar flares which are not normally longer than a few hours \citep[e.g.][]{pye15}. Furthermore, the lack of any stellar counterpart at other wavelengths, particularly in the HST observations of \srciii, make the association with a star in our Galaxy unlikely. These HST observations contained sources down to $m_{\rm F814W}\sim26$, which when applying a distance modulus of 15 corresponding to 10 kpc implies $M_{\rm F814W}\sim11$. This does not rule out a white dwarf or cool main sequence star, however.

\subsection{Sources in the background of these nearby galaxies?}

With the exception of \srcv\ and \srciv, which appear to be in the disks of NGC 7793 and M81 respectively and therefore not likely to be in the background of these galaxies, the other sources may be in the background of the galaxy they appear to be associated with. If so, their timescales and fluxes compare well with tidal disruption events \citep[TDEs, e.g.][]{auchettl17}. Assuming a typical TDE X-ray luminosity of $10^{44}$ \ergs, this puts the sources at $z \sim 0.1$. At this distance, all known TDE host galaxies \citep[e.g.][]{french20} have a $V$-band magnitude of 17.5--21.5. The \hst\ observation of \srciii\ would have detected a background galaxy where the $F606W$ (wide $V$ band) observations reach $m_{\rm F606W}\sim26$. Furthermore, TDEs are usually also bright in the optical/UV, so the lack of a UVOT counterpart also argues against a TDE. Gamma-ray bursts also have much shorter timescales than these X-ray transients, of the order hours \citep[e.g.][]{tarnopolski15}. The after-glows of Gamma-ray bursts are longer, but are usually accompanied by emission at other wavelengths.

\section{A new population of transient ULXs}

If we can rule out foreground sources in our Galaxy, and background sources, we are left with the conclusion that these X-ray sources are associated with the galaxies close in projected separation to them, i.e. NGC 4945, NGC 7793 and M81. At the distances to these galaxies, all at 3--4 Mpc, their peak luminosities are 2--3$\times10^{39}$ \ergs. While supernovae can produce these X-ray luminosities on the timescales observed \citep{chevalier17}, the lack of optical/UV emission disfavors a supernova origin for these sources. This then makes these sources likely ULXs. While these galaxies are known hosts to other ULXs, these other ULXs are relatively persistent sources, whereas the sources we have identified are transient.

With the exception of \srciii, all our sources are within the $D_{\rm 25}$ isophotal ellipses of their apparent host galaxies which are traditionally used for the creation of ULX catalogs \citep[e.g.][]{earnshaw19,walton21}. \srciii\ is 1.7 times further from the center of NGC 7793 than the semi-major axis of that galaxy's isophotal ellipse and therefore would have been missed by these catalogs.

For \srcv, the \swiftxrt\ position was consistent with the nucleus which initially made it a candidate TDE, albeit a low luminosity one. However, the \nustar\ spectrum revealed a turnover in the spectrum of the source that is characteristic of ULXs, and not seen so far in TDEs. Furthermore, the more precise position provided by \chandra\ ruled out the nucleus, confirming that the source was indeed a ULX.

\subsection{A search for the stellar counterparts}
\label{sec_hst}

As mentioned in Section \ref{sec_srcs}, \hst\ has observed the regions of \srciii, \srciv\ and \srcv. \srciii, which is in the outskirts of NGC 7793, was observed as part of the GHOSTS survey \citep{radburnsmith11} with the ACS and F606W and F814W filters. While none of the sources detected in these observations are within the X-ray positional uncertainty region, several are nearby, the properties of which may yield clues as to the environment of the source. The position of the X-ray source is 7.6\arcmin\ from the center of NGC 7793, which implies a projected distance of 8.4 kpc assuming a distance of 3.8 Mpc to the galaxy (Figure \ref{fig_ngc7793_img}). The \hst\ observations were all taken many years before the X-ray transients, so the photometry is unlikely to be contaminated by the accretion disks. 

\begin{figure*}
\begin{center}
\includegraphics[trim=10 20 20 0, width=180mm]{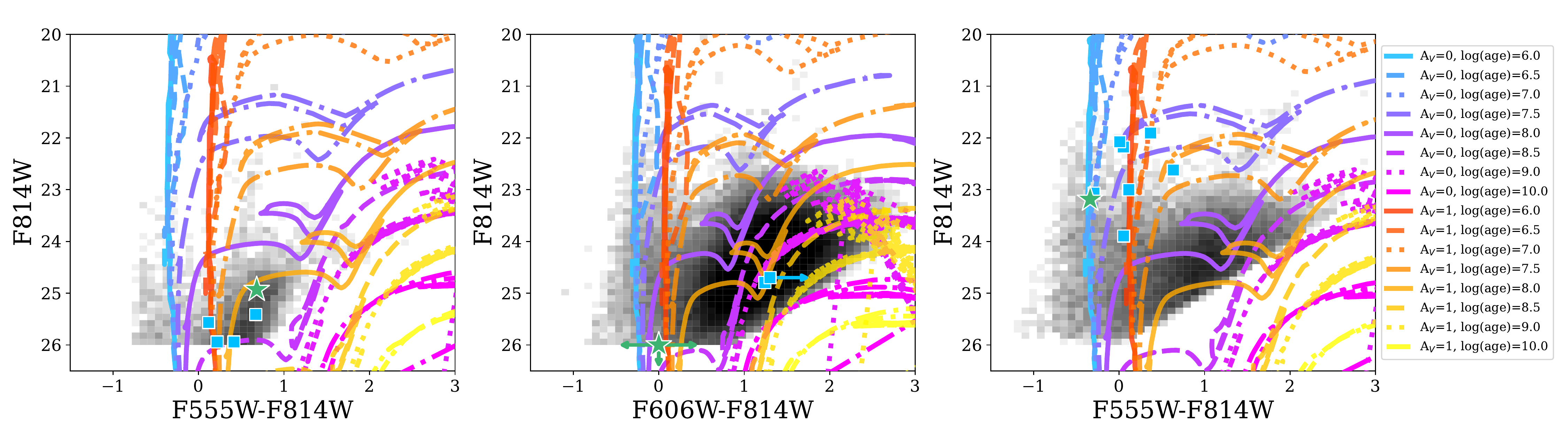}
\end{center}
\caption{ Color-magnitude diagram of HSC sources in and around \srciii\ (left), \srciv\ (middle), and \srcv\ (right). The closest sources are shown with a green star, and the other sources within the positional error circle are shown with blue squares. Arrows indicate upper limits. The lines represent stellar isochrones showing where stars of a certain age are expected to lie.}
\label{fig_cmd}
\end{figure*}

 In Figure \ref{fig_cmd} we present color magnitude diagrams (CMDs) with all stars in the vicinity of the ULX plotted in the black histogram in the background. The green star (or green arrows) on each CMD indicates the star closest to the ULX source position and the blue squares indicate stars within the positional error circle (with the exception of \srciii\ where we show the stars within 10\arcsec). Some stars had non-detections in the HST filters we plot here, and these non-detections are indicated with arrows. In the middle panel, we have two stars plotted with non-detections. The green arrows indicate a star that was not detected in either band plotted, and the star plotted with a horizontal arrow was detected in the F814W band, but not the F606W band.

We include isochrones from the Padova stellar models \citep{marigo2008,bressan2012,marigo2017} at different ages, which are listed in the figure legend. The purple lines represent isochrones of various ages with no dust extinction applied (A$_{V}$=0.0) and the orange lines represent isochrones with 1 magnitude of dust extinction (A$_{V}$) applied using the reddening coefficients from \cite{schlafly11} in the HST filters presented in each CMD.

For \srciii\ the closest stars lie on the red giant branch (RGB) of the CMD described in \cite{radburnsmith11}. The closest source falls in a region identified by \cite{radburnsmith12} as belonging to old RGB stars with ages of 1--10 Gyr. Our isochrones imply they could be 100--300 Myr or 1--30 Myr with 1 A$_{V}$ of extinction. The case is similar for \srciv. For \srcv, the closest star lies on the main sequence with an age of 1--10 Myr. The other stars within the positional error circle may have ages of up to 30 Myr.

\subsection{Previous results on transient ULXs}

While other transient ULXs have previously been presented in the literature, many, if not all of these sources were discovered serendipitously, and not in real time. As far as we know, this is the first study to carry out a systematic search for transient ULXs outside of our Galaxy in real time, allowing us to conduct a more detailed study with follow up observations, such as with \swift\ to get well-sampled lightcurves, \chandra\ to obtain more precise positions, and \nustar\ to obtain a broadband spectrum.

We have reported on two transient ULXs in NGC 4945, however two other transient X-ray sources, Suzaku J1305-4931 and Suzaku J1305-4930, with ULX or close-to-ULX luminosities were reported by \cite{isobe08} and \cite{ide20} from {\it Suzaku} observations of the galaxy. The sources were detected at positions of 13h 05m 05.s5, -49\degree\ 31\arcsec\ 39\arcsec\ and 13h 05m 17s.0, -49\degree\ 30\arcmin\ 15\arcsec\ respectively. Suzaku J1305-4931 was active in 2006 Jan and Suzaku J1305-4930 was active in 2010 July, and lasted less than 6 months. Both {\it Suzaku} sources were close to our \swiftxrt\ sources ($\sim1$\arcmin\ from either), but closer to the plane of the galaxy.

In addition to \srciii\ and \srcv\ in NGC 7793, \cite{quintin21} reported the discovery of another transient ULX in that galaxy found while looking for long-term variability of \xmm\ sources in the 4XMM-DR9 catalogue. The source, which they name NGC 7793 ULX-4, was active for 8 months from 2012 May--Nov. The ULX had a position of 23h 57m 47s.9 -32\degree\ 34\arcmin\ 57\arcsec\ which is close to the center of the galaxy, $\sim8$\arcmin\ from \srciii\ and $\sim40$\arcsec\ from \srcv. They also reported a pulsation signal at 2.52 Hz from the \xmm\ data making it the second ULX pulsar in that galaxy.

Examples of other transient ULXs presented in the literature are ones in M31 \citep{middleton12,middleton13}, M51 \citep{brightman20}, M83 \citep{soria12}, M86 \citep{vanhaaften19}, NGC~55 \citep{robba22}, NGC~300 \citep{carpano18}, NGC~821 \citep{dage21}, NGC~925 \citep{earnshaw20}, NGC~1365 \citep{soria07}, NGC~3628 \citep{strickland01}, NGC~4157 \citep{dage21}, NGC~5907 \citep{pintore18}, NGC~6946 \citep{earnshaw19b} and NGC~7090 \citep{liu19,walton21a}.

One of the best studied transient ULXs is NGC~300~ULX1, which was classified as a supernova imposter in 2010, with an observed X-ray luminosity of 5$\times10^{38}$ \ergs \citep{binder11}. The source was observed at lower fluxes in observations made in 2014 \citep{binder16} but then reached ULX luminosities during observations made in 2016 with \lx$\sim5\times10^{39}$ \ergs\ during which pulsations were detected \citep{carpano18} identifying it as a ULX pulsar powered by a neutron star. Regular \swift\ monitoring of the source in 2018 revealed that the source initially persisted at these luminosities but then went into decline. Spectral analysis showed a hard spectrum.

Another source, Swift~J0243.6+6124, was an X-ray transient found in our own Galaxy, identified by \swift/BAT \citep{cenko17} and with no previously reported activity. The source reached an X-ray luminosity of $2\times10^{39}$ \ergs\ in a period of around 30 days, before steadily declining over a period of $\sim100$ days \citep{wilsonhodge18}. The detection of pulsations also identified it as a neutron star accretor \citep{kennea17} and \cite{kong22} reported the detection of a cyclotron resonance scattering feature at 146 keV with HXMT, allowing for the estimation of the magnetic field strength to be $\sim1.6\times10^{13}$ G. The source exhibited re-brightenings in the X-ray after the decline, albeit to peak luminosities around 2 orders of magnitude less than the initial outburst \citep{vandeneijnden19}. The companion star is a known Be type.

RX~J0209.6-7427 is a Be X-ray binary in the SMC and also briefly became ULX pulsar in 2019 \citep{chandra20,vasilopoulos20b,coe20}. The spin period was 9.3\,s and reached a luminosity of 1--2$\times10^{39}$ \ergs\ similar to the super-Eddington outburst of SMC X-3 \citep{tsygankov17}. \cite{karino22} discussed the possibility that a large number of ULXs are formed of Be HMXBs.

 We also note that the peak luminosities of these sources, disk temperatures, and implied inner disk radii are similar to the brightest outbursts from Galactic X-ray binaries such as GRO J1655-40, GX 339-4 and XTE J1550-564 when considering fits with the disk model.

\subsection{Are these new transient ULXs also Be X-ray binaries?}

As described above, many well known ULX transients are Be X-ray binaries in outburst, therefore it is reasonable to ask if these new systems are also Be XRBs. The peak luminosities of 2--3$\times10^{39}$ \ergs\ are consistent with the type II outbursts from these sources, however Be XRBs typically have longer rise times, up to 50 days from detection to peak, whereas our ULX transients have rise times of 10 days or less where the the lightcurves are well sampled \citep{reig13}. Furthermore, Be stars are young and massive, at odds with the older stellar population that \srciii\ and \srciv\ are found in. For \srcv, the potential stellar counterparts are young and massive, therefore we cannot rule out a Be star in this case.

\subsection{Modeling the lightcurves with a disk-instability model}

A model was presented in \cite{hameury20} to explain the transient ULX phenomenon with a disk instability model previously used to explain dwarf novae and other X-ray transients. There, the super-Eddington outbursts can be explained by thermal-viscous instabilities in large unstable disks with outer radii greater than $10^{7}$\,km. They showed that this model can successfully reproduce the lightcurve of the transient ULX M51 XT-1 presented in \cite{brightman20}, with derived accretion rates of 6--15$\times10^{19}$ g\,s$^{-1}$ depending on the mass of the accretor. 

We fit the observed transient ULX lightcurves using these models. \cite{hameury20} provides analytical formulas that accurately approximate the observable properties of the outbursts. According to this model, the accretion disk in outburst is brought into a fully hot state, then the disk mass decreases until the surface density at the outer edge of the disk becomes inferior to the critical value below which quasi--stationary hot states can no longer exist. This results in a cooling wave, propagating into the disk from its outer edge, bringing the whole disk back into a low state.
When the disk is fully in the hot state, it is close to steady state with a mass accretion rate that is constant with radius and larger than the mass transfer rate from the secondary. \citet{hameury20}, following \citet{rk01} have shown that during this phase, the accretion rate \textsl{does not} decrease exponentially, but according to:
\begin{equation}
\dot{M}=\dot{M}_{\rm max} \left[ 1+ \frac{t}{t_0} \right]^{-10/3},
\label{eq:phase1}
\end{equation}
where $\dot{M}=\dot{M}_{\rm max}$ is the initial mass accretion rate and $t_0$ is a characteristic timescale which depends on $\dot{M}_{\rm max}$ and the disk parameters (size, viscosity):
\begin{equation}
t_0 = 3.19 \alpha_{\rm 0.2}^{-4/5} M_1^{1/4} r_{12}^{5/4} \dot{M}_{\rm max,19}^{-3/10} \; \rm yr,
\label{eq:t0}
\end{equation}
where {$\dot{M}_{\rm max,19}$ is $\dot{M}_{\rm max}$ measured in units of $10^{19}$~g~s$^{-1}$,} $r_{12}$ is the disk size in units of 10$^{12}$ cm, $M_1$ the accretor mass in solar units and $\alpha_{0.2}$ the Shakura-Sunyaev viscosity parameter normalized to 0.2. Therefore, for given binary parameters and disk viscosity, the initial time evolution of the disk depends only on one free parameter, the initial accretion rate. Conversely, the determination of $t_0$ from observations (at time $t=t_0$, the mass accretion rate is one tenth of its initial value) enables one to determine the disk size. This phase lasts until the accretion rate falls below the critical rate $M_{\rm crit}^+$ at which the hot solution can no longer exist at the outer radius. Using Eqs. (\ref{eq:phase1}), (\ref{eq:t0}) and the fits for $M_{\rm crit}^+$ provided by \citet{hameury20} in their Eq. (9), one finds that the duration of this phase is:
\begin{equation}
\Delta t_1 = t_0 [ 1.38 t_0^{-0.50} M_1 ^{0.25} \dot M_{19,\rm max}^{0.15} f_{\rm irr}^{0.15} \alpha_{0.2}^{-0.4} -1]
\label{eq:duration}
\end{equation}
and is followed by a rapid decay phase during which the accretion rate drops steeply; the duration of this final phase is 
\begin{equation}
   \Delta t_2 = 0.74 \; M_1^{0.37} f_{\rm irr}^{0.15} \alpha_{0.2}^{-0.8} r_{12}^{0.62} \; \rm yr. 
    \label{eq:end}
\end{equation}
where $f_{\rm irr} \sim 1$ is a parameter describing the effect of irradiation on the disk.

The above equations describe the time evolution of the mass accretion rate as a function of only four parameters: the initial (i.e. peak) mass accretion rate, $t_0$, $M_1$ and $\alpha_{0.2}$ to which one could add $f_{\rm irr}$ which enters only via $f_{\rm irr}^{0.15}$. The parameters $\Psi$ and $\xi$, as defined in \citet{hameury20}, were taken equal to 1.3 and 6.3 respectively, since, as discussed in \citet{hameury20}, these values provide the best agreement between the numerical results and their analytical approximations. $\Psi$ accounts for deviations of the opacities from the Kramers' law and $\xi$ is the ratio between the rate at which the inner, hot disk mass decreases and the accretion rate at the inner edge; it is larger than unity because of the strong mass outflow at the cooling front.

In order to compare the model predictions with observations, one must convert mass accretion rates into luminosities; we used $L_{\rm X} = (1 + \ln \dot m) [1+\dot m^2/b] L_{\rm Edd}$ for luminosities larger than the Eddington value, where $b$ is a constant which differs from, but is related to the beaming parameter \citep[see][]{king09}. This relation differs somewhat from the formula derived by \citet{king09}, valid only for strong beaming; we modified it in order to account for a smooth transition with the case where beaming is negligible, as explained in \citet{hameury20}.

If the final decay is not observed, one cannot determine the viscosity parameter, since the disk radius, which enters Eq. (\ref{eq:t0}), is not known a priori. One can nevertheless obtain upper limits on $\alpha_{0.2}$ because the observed duration of the outbursts sets a lower limit on $\Delta t_1$. Moreover, one  expects significant degeneracies between $t_0$, $\dot M_{\rm max}$ and $M_1$ when the light curve does not deviate much from an exponential (i.e. when $t$ never becomes larger than $t_0$) which is defined by two parameters only. On the other hand, useful constraints can be obtained when the final decay is observed, and therefore the duration of the hot phase is measured. Because of the rapid drop off during the final decay, one does not expect to get significant constraints from the shape of this phase, but one at least gets a determination of $\Delta t_1$, and some constraint, usually in form of an upper limit, on $\Delta t_2$. This basically sets three strong constraints on four parameters, implying that degeneracies should still exist that preclude the simultaneous determination of $M_1$ and $\alpha_{0.2}$ unless observational uncertainties are very low.

\begin{table*}
\centering
\caption{Fits of the outburst light curves}
\label{tab_fits}
\begin{center}
\begin{tabular}{c c c c c c c c c c}
\hline      &       & \multicolumn{4}{c}{\texttt{powerlaw} model} & \multicolumn{4}{c}{\texttt{diskbb} model} \\

Source name	& $M_1$ & $t_0$ & $\dot M_{\rm max}/\dot M_{\rm Edd}$	& $\alpha$	& $\chi^2$/DOF & $t_0$ & $\dot M_{\rm max}/\dot M_{\rm Edd}$	& $\alpha$	& $\chi^2$/DOF    \\
 			& 	    & (days)&			 							& 			& 			   & (days)  \\
\hline
\multirow{2}{*}{Swift J130456.1-493158} &  1.4  &  159.8 &  18.67 &  $< 1.4$ &   0.90/6 &   78.4&   8.97&  $<1$   &   0.61/6 \\
										&  10   &   74.9 &   4.23 &  $< 7  $ &   0.86/6 &   96.0&   0.89&  $<3$   &   0.79/6 \\
\multirow{2}{*}{Swift J130511.5-422933} &  1.4  &  395.2 &  20.89 &  0.37    &  19.06/21&  309.6&  14.93&  0.33   &  15.55/21 \\
										&  10   &  164.2 &   5.33 &  1.39    &  18.52/21&  199.3&   2.20&  1.06   &  18.51/21 \\
\multirow{2}{*}{2SXPS J235825.7-323609} &  1.4  &  365.0 &  13.75 &  0.35    &  15.78/12&  271.6&  10.07&  0.33   &  12.34/12 \\
										& 10    &  335.3 &   1.56 &  1.06    &  18.02/12&  322.9&   0.96&  0.90   &  16.63/12 \\	
\multirow{2}{*}{Swift J235749.9-323526} &  1.4  &  216.4 &  19.49 &  $<1  $  &  10.71/11&  184.9&  15.67&  $<0.9$ &   9.62/12\\
                                        & 10    &   89.4 &   4.72 &  $<6$    &  14.72/11&  102.5&   2.45&  $<4$   &  14.60/12\\
\multirow{2}{*}{Swift J095520.7+690401} &  1.4  &   61.6 &   9.98 &  $<0.5$  &   4.59/5 &   65.6&   4.83&  $<0.4$ &   5.80/5\\
										& 10    &   81.2 &   0.90 &  $<1.5$  &   5.36/5 &   81.2&   0.49&  $<1$   &   5.36/5\\
\hline		
\end{tabular}
\end{center}
\end{table*}

Table \ref{tab_fits} shows the results of the fitting procedure,  both when using the {\tt powerlaw} model to convert from count rate to flux, and the {\tt diskbb} model. As expected, the viscosity parameter $\alpha$ can be determined only when the final decay has been observed, i.e. in the case of 2SXPS J235825.7-323609 and Swift J130511.5-422933. In this case, the value of $\alpha$ we obtain is large, and typically much larger than the value $\alpha \sim 0.1 - 0.2$  determined when fitting the light curve of cataclysmic variables \citep{s99,kl12}. This should not come as a surprise since \citet{tlh18} found that much larger values of $\alpha$, in the range 0.2 -- 1 are obtained when considering low-mass X-ray binaries; they attributed this large value of the viscosity parameter to the existence of strong winds in these system that carry away matter and also angular momentum. We take $b=73$, as in \citet{king09}, but because of the relatively large size of the error bars, the fits are not sensitive to the value of the beaming parameter. For M51 XT-1, with two data points with relatively small error bars, \cite{hameury20} were able to exclude $b=20$, but found that $b=200$ gives an acceptable fit.

We also note that, again as expected, the primary mass cannot be determined from fitting the observed light curves. The fits are equally good for neutron star and for 10~M$_\odot$ black hole accretors. The only difference is that the lower $M_1$, the shorter the duration $\Delta t_1$ of the main outburst phase, and long outbursts may require unrealistically low values of $\alpha$ in the neutron star case.

The maximum accretion rate is in all cases is close to or larger than the Eddington limit, except in the case of Swift J095520.7+690401 for large primary masses.

\begin{figure}[h]
\begin{center}
\includegraphics[trim=10 20 20 0, width=90mm]{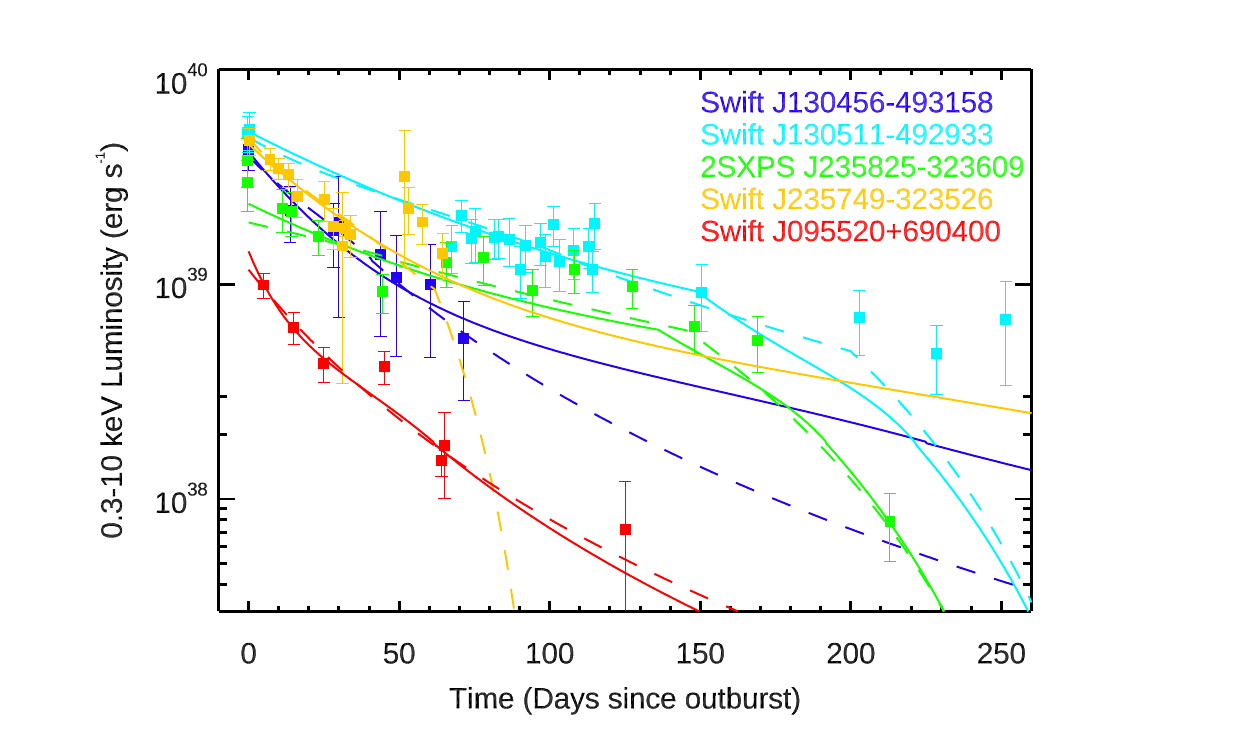}
\end{center}
\caption{Lightcurves of all the transients presented here fitted with the disk instability model presented by \cite{hameury20}.  Luminosities are from the {\tt powerlaw} spectral model. Upper limits are omitted in the plot for clarity. The solid lines represent the model assuming a 1.4 \msol\ accretor, whereas the dashed lines represents a 10 \msol\ accretor.}
\label{fig_model_fits}
\end{figure}

The fit quality, as measured by the $\chi^2$ compared to the number of degrees of freedom is quite good in all cases, except for 2SXPS J235825.7-323609. The reason for this is the existence of a low data point at $t \sim 50$~d, and, to a lesser extent, a slightly steeper final decay than predicted by the model.  Although the $\chi^2$ value is good for Swift J130511.5-422933, the observed drop between $t=250$~d and $t=284$~d is too sharp to be accounted for by the model. The detection at $t \sim 284$~d is somewhat marginal and the estimate of the X-ray luminosity becomes questionable because of uncertainties in the spectral model; it is however unlikely that the current model can reproduce the sharp cutoff observed in this source. This would mean mean that the cooling front propagates faster than expected when the propagation is controlled by irradiation with a constant efficiency. Dropping this hypothesis might solve this problem, at the expense of a new and uncontrolled parameter; given other oversimplifying assumptions of the model, notably about winds, this would be of limited interest. One should note that similar problems are encountered when modeling outbursts of sub-Eddington X-ray transients \citep{tdl08}. Although the $\chi^2$ value is also good for Swift J235749.9-323526, the model does not reproduce what appears to be a plateau or rebrightening at $t \sim 50$~d which is unlikely due to changes in the accretion disk. The short drop at $t \sim 10$~d is also not accounted for by the model, and the {\it NuSTAR} data point has not been included in the fit. We show the fits to all sources in Figure \ref{fig_model_fits}.

\subsection{Implications for the wider ULX population}

We summarize the properties of the sources in Table \ref{table_prop}. We find that the average \nh$=5.7\times10^{21}$ \cmsq\ with a standard deviation of $3.8\times10^{21}$ \cmsq. The average $\Gamma$ is 2.3 with a standard deviation of 0.4. This is consistent within the standard deviations of the sample of persistent sources from \cite{gladstone09}, where the average \nh$=2.8\times10^{21}$ \cmsq\ with a standard deviation of $1.7\times10^{21}$ \cmsq\ and the average $\Gamma$ is 2.3 with a standard deviation of 0.5. Therefore we do not see any significant spectral differences between our transient sources and their persistent counterparts. For \srcv\ where we obtain \nustar\ data to extend the spectral coverage to higher energies, the {\tt diskpbb} model was preferred over the power-law one, typical of ULX spectra as shown by \cite{walton18c}. Again, the parameters of this model were consistent with those seen in the persistent sources.

\begin{table*}
\centering
\caption{Summary of source properties}
\label{table_prop}
\begin{center}
\begin{tabular}{c c c c c c c c}
\hline
Source name	& Host galaxy 	& Best position 	& 			& uncertainty 	& \nh 	& $\Gamma$ 	& \lx\ (peak) \\
			& 			& RA (\degree)	& Dec (\degree)& (\arcsec) 	& (\cmsq) 	& 			& (\ergs) \\
\hline
\srci\ & NGC 4945 & 196.23479 & -49.53324 & 0.53 & $1.1^{+0.5}_{-0.4}\times10^{22}$ & $2.8^{+0.6}_{-0.5}$ & $2\times10^{39}$ \\
\srcii\ & NGC 4945 & 196.2985 & -49.4928 & 2.4 & $6.7^{+2.1}_{-1.7}\times10^{21}$ & $2.2\pm0.3$ & $2\times10^{39}$ \\
\srciii\ & NGC 7793 & 359.60828 & -32.60291 & 1.0 & $2.0^{+1.2}_{-1.0}\times10^{21}$ & $2.0\pm0.3$ & $3\times10^{39}$ \\
\srcv\ & NGC 7793 & 359.45793 & -32.59110 & 0.57 & $2.1^{+0.9}_{-0.8}\times10^{21}$ & $2.0\pm0.3$ & $3\times10^{39}$ \\
\srciv\ & M81 & 148.83697 & +69.06737 & 0.33 & $<6.8\times10^{21}$ & $2.6^{+1.6}_{-1.1}$ & $2\times10^{39}$ \\\hline
\end{tabular}
\end{center}
\end{table*}

We have found that two of our sources appear to lie in a population of old RGB stars. Interestingly, \cite{wiktorowicz17} predicted that the majority of neutron star ULXs have low-mass ($<$1.5 \msol), red giant donors. According to \cite{wiktorowicz17} red-giant donor NS-ULXs form at late times, and start with the primary becoming a Oxygen-Neon white dwarf. When the secondary becomes a red giant and fills its Roche lobe, the primary accretes additional mass and forms a NS due to an accretion induced collapse (AIC). Following this, the RG refills its RL and a short ($0.1<\Delta t<0.2$ Myr) ULX phase occurs. We have not unambiguously identified the donor stars of these sources as red giants, and neither do we know they are neutron stars, but these properties do match well, albeit the timescales are much shorter than suggested by \cite{wiktorowicz17}.

It has been suggested that fast radio bursts (FRBs) may be associated with ULXs \citep{sridhar21}. In this model, the accreting compact object is a black hole or a non-magnetar neutron star as in \cite{king09}. One FRB with an intriguing similarity to our transient ULXs is FRB 20200120E which was found in the outskirts of M81 \citep{bhardwaj21}. The FRB was localized to a globular cluster with an old stellar population which challenged the magnetar models that invoke young magnetars formed in a core-collapse supernova but would be consistent with the \citep{sridhar21} scenario. AIC of a white dwarf was also suggested as a possible formation channel \citep{kirsten22}. However, to date, no FRB has been reported at the position of a ULX.

In addition to the 5 transient ULXs we have presented here, 3 further transient ULXs were serendipitously discovered in the same galaxies from previous observations implying that the rates of these sources is potentially high. While our sample of 5 sources is small, we next attempt to estimate the rates of transient ULXs in these galaxies, and compare these to their persistent counterparts. 

For NGC 4945 we found 2 transient ULXs in searches of observations over 3.0 years from 2019 Dec to 2022 Dec, implying a rate of 0.7$\pm0.5$ year$^{-1}$. Using the same technique to identify the transient sources, and in the same period, we found 4 persistent sources classified as ULXs identified in SIMBAD as [CHP2004] J130518.5-492824, [BWC2008] U31, [CHP2004] J130521.2-492741 and [BWC2008] U32. 

For NGC 7793 we found 2 transient ULX in searches of observations over 5.0 years from 2017 Dec to 2022 Dec, implying a rate of 0.4$\pm0.3$ year$^{-1}$. In the same period, we found 2 persistent sources classified as ULXs, P9 and P13. 

For M81 we found 1 transient ULX in searches of observations over 9 months from 2022 Apr to 2022 Dec, implying a rate of 1.3 year$^{-1}$. In the same period, we found 1 persistent source classified as a ULX, [LB2005] NGC 3031 ULX2. 

 If we compare the number of transient ULXs in any one snapshot to the number of persistent ULXs, as would be done when computing the X-ray luminosity function of a galaxy \citep[e.g.][]{lehmer19}, the persistent sources would dominate the high end. However, if we take the total number of sources that have exceeded $10^{39}$ \ergs\ over the time period of our searches, the transient ULX numbers roughly equal those of the persistent ones. Further, if we integrate the derived transient ULX rates over the timescales that the persistent source have been detected, several decades, then the transient ULX numbers would dominate the persistent ones. In other words, the number of systems that exhibit ULXs luminosities in each of these galaxies is dominated by transients rather than persistent sources.

Since we have only considered galaxies where a transient ULX has been identified in our searches, we cannot extend this conclusion to all galaxies. The rates are also biased by the \swift\ targeting and our incomplete search of observations. A more systematic search using \erosita\ data could reveal the true rate. However, we note that the 6-month scanning pattern of \erosita\ means some of the sources we identified can be missed.

While the duration of the transient sources studied here is well determined, the duration of the persistent sources is not well known. However evidence points to their far longer duration. For example, the collisionally ionized bubbles surrounding Holmberg IX X-1, NGC 1313 X-2, NGC 7793 S26 and NGC 5585 ULX have estimated dynamical ages of $\sim10^{5}$ years \citep{pakull02,pakull10,moon11,weng14,soria21}. 

\section{Summary and Conclusions}

We have presented results on five newly found X-ray transients in the fields of nearby galaxies identified in a search of \swiftxrt\ observations. Our results are as follows:

\begin{itemize}
\item The timescales (60--400 days), fluxes ($\sim10^{-12}$ \ergcms), and lack of bright optical/UV counterparts argue against foreground sources in our Galaxy such as stars or X-ray binaries, and more distant sources such as tidal disruption events or Gamma-ray bursts.
\item These X-ray transients appear to be ultraluminous X-ray sources associated with the nearby galaxies of NGC 4945, NGC 7793 and M81 with peak luminosities of 2--3$\times10^{39}$ \ergs.
\item For 4 out of 5 sources, modeling the lightcurves of these transients with the disk instability model of \cite{hameury20} implies that the mass accretion rate through the disk is greater than the Eddington rate regardless of whether a 1.4 \msol\ neutron star or 10 \msol\ black hole is assumed.
\item For the three sources where {\it HST} imaging enables a search for a stellar counterpart.  We plotted CMDs with stellar isochrones which imply varying ages of the potential stellar counterparts.
\item The rate of transient ULXs for these three galaxies is in the range of 0.4--1.3 year$^{-1}$.  While persistent ULXs dominate the high end of galaxy luminosity functions, the number of systems that produce ULX luminosities are likely dominated by transient sources.
\item The potential dominance of transient ULXs may imply results on ULXs may be biased by studies of persistent sources.
\end{itemize}

\facilities{Swift (XRT, UVOT), CXO, NuSTAR, XMM, VLA} 

\software{{\tt CIAO} \citep{fruscione06}, {\tt XSPEC} \citep{arnaud96}}

\acknowledgements{We thank the anonymous referee for the careful review of our manuscript, and their helpful comments which improved its quaility.

We wish to thank the \swift\ PI, Brad Cenko, for approving the target of opportunity requests we made to observe these transient sources, as well as the rest of the \swift\ team for carrying them out. We also acknowledge the use of public data from the \swift\ data archive. This work made use of data supplied by the UK Swift Science Data Centre at the University of Leicester. 

We also wish to thank Patrick Slane, Director of the Chandra X-ray Center, for approving the DDT requests to observe \srci, \srciv\ and \srcv, and the \chandra\ team for carrying out the observations.

In addition we wish to thank the \nustar\ PI, Fiona Harrison, for approving the DDT request we made to observe \srcv\ as well as the \nustar\ SOC for carrying out the observation. This work was also supported under NASA Contract No. NNG08FD60C. \nustar\ is a project led by the California Institute of Technology, managed by the Jet Propulsion Laboratory, and funded by the National Aeronautics and Space Administration. This research has made use of the NuSTAR Data Analysis Software (NuSTARDAS) jointly developed by the ASI Science Data Center (ASDC, Italy) and the California Institute of Technology (USA).

This research has made use of data obtained from the Chandra Source Catalog, provided by the Chandra X-ray Center (CXC) as part of the Chandra Data Archive.

This work was also based on observations obtained with XMM-Newton, an ESA science mission with instruments and contributions directly funded by ESA Member States and NASA

This research has made use of data and/or software provided by the High Energy Astrophysics Science Archive Research Center (HEASARC), which is a service of the Astrophysics Science Division at NASA/GSFC.

Based on observations made with the NASA/ESA Hubble Space Telescope, and obtained from the Hubble Legacy Archive, which is a collaboration between the Space Telescope Science Institute (STScI/NASA), the Space Telescope European Coordinating Facility (ST-ECF/ESAC/ESA) and the Canadian Astronomy Data Centre (CADC/NRC/CSA).  The Hubble Source Catalog can be accessed via \dataset[DOI]{https://doi.org/10.17909/T97P46}, and the specific observations used can be accessed via \dataset[DOI]{https://doi.org/10.17909/jmva-dc49}.

The National Radio Astronomy Observatory is a facility of the National Science Foundation operated under cooperative agreement by Associated Universities, Inc.

JMC's research was supported by an appointment to the NASA Postdoctoral Program at the NASA Goddard Space Flight Center, administered by Oak Ridge Associated Universities under contract with NASA. MH acknowledges support from an ESO fellowship and JPL was supported in part by a grant from the French Space Agency CNES. 
}

\bibliography{Swift XRT transients paper.bbl}

\end{document}